%% file: GAFA.tex
\documentclass[sigplan,10pt,screen]{acmart}
\renewcommand\footnotetextcopyrightpermission[1]{}
\usepackage{tikz}
\usepackage{amsmath}
\usepackage{graphicx}
\usepackage{graphics}
\usepackage{float}
\usepackage{subfigure}
\usepackage{multirow}
\usepackage{color}
\usepackage{tabularray}
\usepackage{kantlipsum}
\usepackage{soul}
\usepackage{utfsym}
\usepackage{booktabs}
\usepackage{enumitem}
\usepackage{url}
\usepackage{tabularx}
\usepackage{ragged2e}
\usepackage{makecell}
\usepackage{subcaption}
\usepackage{caption}

\definecolor{fcolor}{cmyk}{0.55,1,0,0.15}
\definecolor{darkred}{RGB}{192,0,0}

\newcommand*\redcircled[1]{\tikz[baseline=(char.base)]{
            \node[shape=circle,fill=red,font=\bfseries,inner sep=0.3pt] (char) {\textcolor{white}{#1}};}}

\iftrue
\newcommand{\yi}[1]{{#1}}
\newcommand{\jie}[1]{{\color{red}[\textbf{\sc jie}: \textit{#1}]}}

\else
\newcommand{\yi}[1]{}
\newcommand{\jie}[1]{}
\fi

\begin{document}
\begin{sloppypar}

\title[GNStor]{
GNStor: Design of GPU-Native High-Performance\\Remote All-Flash Array 
}

\author{Shushu Yi}
\affiliation{%
 \institution{Peking University}
 \country{}}

\author{Wenbo Wu}
\affiliation{%
 \institution{Peking University}
 \country{}}

\author{Guoci Chen}
\affiliation{%
 \institution{Peking University}
 \country{}}

\author{Junrong Zhu}
\affiliation{%
 \institution{Peking University}
 \country{}}

\author{Shengwen Liang}
\affiliation{%
 \institution{ICT, CAS}
 \country{}}

\author{Mao Bo}
\affiliation{%
 \institution{Xiamen University}
 \country{}}

\author{Chenying Huan}
\affiliation{%
 \institution{Nanjing University}
 \country{}}

\author{Chen Tian}
\affiliation{%
 \institution{Nanjing University}
 \country{}}

\author{Jie Zhang}
\affiliation{%
 \institution{Peking University}
 \country{}}

\begin{abstract}
GPU has become the leading computing device for a wide range of data-intensive applications, which tightly collaborates with remote all-flash array (AFA) to accommodate ever-expanding datasets, facilitate multi-client data sharing, and guarantee fault tolerance.
Although GPU is the center of computation, all I/O processes in existing GPU-AFA systems are still CPU-centric. CPU orchestrates remote I/O requests and executes a centralized AFA engine to take charge of AFA-level functionalities (e.g., access control and metadata persistence). This design disparity suffers from substantial CPU-GPU interaction overhead and I/O traffic amplification, compromising end-to-end I/O performance.

In this work, we present \emph{GNStor}, a GPU-native AFA system that enables GPU to directly access remote AFA without CPU intervention in the I/O path, thereby fully exploiting the performance of AFA.
Specifically, GNStor first proposes a GPU-centric NVMe over RDMA (NoR) software stack (named \emph{GNoR}), paving a fast path for GPUs to directly initiate NoR I/O requests to SSDs within remote AFA.
GNoR employs an atomic-operation-based I/O orchestration design and follows the single-instruction-multiple-thread (SIMT) execution model of GPU, fully exploiting the massive parallelism of GPU architectures.
To facilitate essential AFA functionalities in a CPU-bypass I/O path, GNStor further designs \emph{deEngine}, a decentralized AFA engine that seamlessly decomposes and integrates AFA-level tasks into each SSD firmware, thereby achieving efficient AFA access at low cost.
Evaluation results show that GNStor achieves 3.2$\times$ higher I/O throughput and reduces application execution time by 31.1\%, compared to state-of-the-art AFA systems.

\end{abstract}

\settopmatter{printfolios=true}
\maketitle
\pagestyle{plain}

\section{Introduction}
\label{sec:intro}
\input{sections/intro}

\section{Background and Challenge}
\label{sec:background}
\input{sections/background}

\section{GNStor Overview}
\label{sec:overview}
\input{sections/overview}

\section{Design and Implementation}
\label{sec:design}
\input{sections/design}

\section{Evaluation}
\label{sec:evaluation}
\input{sections/evaluation}

\section{Related Work and Discussion}
\label{sec:relatedwork}
\input{sections/relatedwork}

\section{Conclusion}
\label{sec:conclusion}
\input{sections/conclusion}

\bibliographystyle{plain}
\bibliography{ref}

\end{sloppypar}
\end{document}

%% file: sections/intro.tex
The last decade has witnessed GPUs emerging as the dominant computing platform to accelerate a wide range of data-intensive applications, such as graph analytics \cite{nguyen2013lightweight,wang2017gunrock,pan2017multi}, recommender systems \cite{resnick1997recommender,burke2011recommender,wang2022merlin}, and large language models (LLMs) \cite{achiam2023gpt,liu2024deepseek,team2023gemini}.
These applications heavily rely on the remote all-flash array (AFA) to accommodate ever-growing datasets, facilitate efficient multi-client data sharing, and provide fault tolerance for long-term data.
For instance, the training corpus of LLMs has reached PB scale,  which is persistently stored on remote AFA and shared across multiple GPU clients \cite{SCADA1,qiu2025geminifs,du2024survey,zhuang2023toolqa}. 
Moreover, LLM training frequently generates checkpoints (including intermediate model weights and optimizer states), which can amount to tens of TBs  \cite{zeng2026gpu,lee2019gpu,yang2024demand}. These checkpoints are periodically written back to AFA with replication (e.g., three replicas in DeepSeek 3FS \cite{dsfs}), ensuring data durability and reliability.

\begin{figure}
 \centering
    \includegraphics[width=1\linewidth]{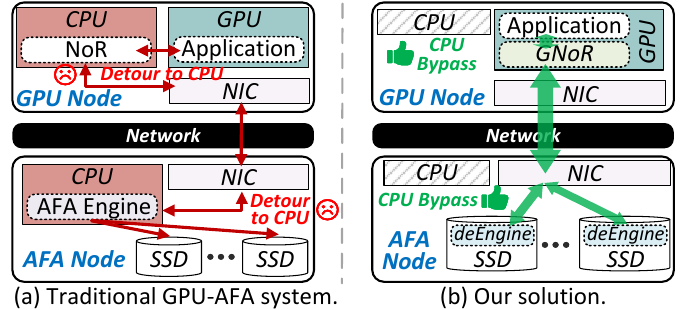}
    \vspace{-20pt}
    \caption{I/O path comparison of different GPU-AFA systems. \label{fig:intro}}
    \vspace{-25pt}
\end{figure}

Despite GPU plays the central role in computation, current practice still relies on CPU to coordinate I/O access to remote AFA. 
As shown in Figure \ref{fig:intro}\textcolor{fcolor}{a}, on the GPU node, CPU is responsible for issuing I/O requests to AFA via remote storage protocols (e.g., \emph{NVMe over RDMA, NoR}), retrieving data from remote AFA to host memory via NIC, and then forwarding it to GPU memory for computation (or vice versa).
Although NVIDIA GPUDirect \cite{gpudirect} enables direct data transfers between GPUs and NICs, it still depends on the CPU for I/O orchestration.
On the storage node (i.e., AFA), the AFA engine also runs on the CPU. 
It manages multiple SSDs, exposes logical volumes to clients in GPU nodes, enables data sharing through access control, enforces data durability via redundancy (e.g., replication), and guarantees crash consistency by properly maintaining and persisting metadata.
These CPU-mediated designs suffer from a detour in I/O path, leading to substantial CPU-GPU interaction overhead and I/O traffic amplification, thereby significantly constraining end-to-end I/O performance. 
Consequently, the performance of remote AFA access from GPU is constrained (e.g., up to 79.1\% throughput degradation in our tests, cf. $\S$ \ref{sec:evaluation}).

One of the promising solutions to tackle this issue is CPU bypass techniques \cite{bam,libnvm,qiu2025geminifs,gofs,sun2025scalio,li2023rubbledb}.
On the GPU node, extensive studies focus on constructing GPU-centric storage stacks \cite{bam,libnvm,qiu2025geminifs,gofs}. 
For example, libnvm \cite{libnvm} and BaM \cite{bam} map the doorbell registers of NVMe SSDs into the GPU address space, enabling GPUs to directly submit NVMe I/O commands to local SSDs.
On the AFA node, modern NICs typically support NoR target offloading mechanism \cite{sun2025scalio,li2023rubbledb}, which employs the \emph{host channel adapter (HCA)} hardware to directly parse NoR commands and interact with SSDs, without trapping into CPU-based software stack.
These approaches succeed in reducing CPU intervention, thereby improving I/O performance in certain scenarios.

Unfortunately, adopting the CPU-bypass design principle to GPU-AFA systems remains challenging.
We summarize the prominent obstacles as follows:

\noindent $\bullet$ \emph{Lack of GPU-centric remote storage software stack:}
Existing GPU-centric solutions \cite{libnvm,bam} are primarily designed for local storage access (i.e., NVMe over PCIe), where the software stack is relatively simple. 
In contrast, enabling remote access requires introducing an additional network layer (e.g., typically RDMA in NoR for high performance \cite{sun2025scalio}), which severely increases system complexity. 
For example, RDMA requires time-consuming memory region (MR) registration to authorize NIC to access physical memory of I/O buffers.
Moreover, due to the fundamental differences between CPU and GPU programming paradigms, lock-based NoR implementation in the CPU-oriented manner cannot be directly ported to GPU.
NoR I/O requests must be carefully orchestrated to ensure extreme parallelism, avoiding resource contention and excessive synchronization overhead even when thousands of GPU threads initiate requests concurrently.

\noindent $\bullet$ \emph{Lack of decentralized AFA engine in CPU-bypass system:}
In an ideal CPU-bypass architecture, GPU-initiated I/O requests can be directly delivered to each SSD in AFA via NIC's HCA, sidestepping the traditional CPU-resident AFA engine. 
While this design shortens I/O path, it also removes the locus for executing I/O-critical AFA functionalities, such as access control and metadata maintenance. 
Without proper coordination from the centralized AFA engine, concurrent accesses from multiple GPU clients can easily compromise system correctness (e.g., two clients may mistakenly occupy the same storage space).
A naive approach is to assign all these tasks to GPU clients. However, this would require each GPU client to frequently coordinate with others for access control, as well as properly maintain and persist metadata for each I/O request. 
Such a design introduces substantial software overhead, which can easily become a scalability bottleneck given high concurrency of GPU scenarios (e.g., 72.0\% throughput degradation in our evaluation).

Tackling these challenges, we introduce \textbf{GNStor}, a GPU-native AFA system that enables GPU to directly access remote AFA without CPU intervention in the I/O path, thereby maximizing end-to-end I/O performance (cf. Figure \ref{fig:intro}\textcolor{fcolor}{b}).
To facilitate GPU-centric access to each SSD in remote AFA, GNStor first designs \emph{GNoR}, a high-performance GPU-oriented remote storage software stack.
The key component of GNoR is a scalable GPU-based NoR driver designed from the ground up to align with the programming paradigms of GPU, fully exploiting its massive parallelism. 
Specifically, rather than directly porting CPU-based NoR implementations, GNoR re-architects the NoR stack by carefully partitioning I/O functionalities. Non-I/O-critical tasks and data structures (e.g., keep-alive handling and NVMe admin queues) remain on the CPU, while I/O-critical components (e.g., NVMe I/O queues and RDMA queues) are encapsulated into \emph{channels} and migrated to the GPU, serving as the basic concurrency abstraction for GPU kernels to submit NoR I/O commands in parallel.
Within each channel, GNoR replaces conventional heavyweight locks with lightweight atomic operations to facilitate scalable concurrent access across hundreds of threads within a GPU kernel.
To mitigate the high overhead of RDMA MR registration \cite{wang2025finemem,cai2018efficient}, GNoR pre-allocates and pre-registers a memory pool for each channel, and manages it with a GPU-friendly multi-level memory allocator. This allocator strives to allocate contiguous memory spaces to reduce the number of RDMA operations required per NoR I/O.

GNoR paves an expressway for point-to-point direct I/O between GPU and remote SSD via NIC. 
Building on this foundation, GNStor further proposes a decentralized AFA engine (named \emph{deEngine}) that supports multiple clients to use multiple AFA volumes (i.e., many-to-many) without invoking CPU in the I/O path.
Specifically, deEngine retains non-I/O-critical AFA management tasks (e.g., volume creation and teardown) on the CPU of AFA node, as they are infrequent and do not impact steady-state performance.
For I/O-critical functionalities (e.g., access control and metadata maintenance), our key insight is that these tasks are substantially similar to the responsibilities already handled by SSD firmware. 
For example, AFA engine verifies whether a client I/O targets a volume with permissions, while SSD firmware similarly validates the legitimacy of I/O to prevent errors (e.g., accessing an illegal address). Moreover, AFA engine maintains mappings from volume-level addresses to SSD addresses, while SSD firmware maintains mappings from SSD logical addresses to physical locations.  
Therefore, deEngine proposes to decompose AFA functionalities and integrate them into each SSD firmware, enabling these tasks to be completed without incurring extra overhead.
For better practicality, GNStor also exposes expressive APIs and libraries for GPU programs, supporting synchronous and asynchronous I/O as well as batched operations.
Comprehensive evaluation results demonstrate that GNStor outperforms leading GPU-AFA system solutions, delivering 3.2$\times$ higher I/O throughput and reducing application execution time by 31.1\%.

\begin{table*}
\renewcommand{\arraystretch}{0.9}
\resizebox{\linewidth}{!}{%
\setlength{\tabcolsep}{5pt}
\centering

\begin{tabular}{|l|l|l|l|l|l|l|}
\hline
\textbf{Workload} & \textbf{Data type} & \textbf{Data size} & \textbf{Access granularity} & \textbf{Retention} & \textbf{Sharing} & \textbf{Performance} \\ \hline

\multirow{2}{*}{Data pre-processing \cite{imagenet100}} 
& Raw samples 
& $10^2$ GB -- $10^1$ PB  
& 1 MB -- $10^1$ GB 
& Years 
& Yes 
& Throughput \\ \cline{2-7}

& Intermediate tensors 
& $10^1$ GB -- $10^2$ TB 
& 1 MB -- $10^1$ GB 
& Minutes 
& No 
& Latency \\ \hline

\multirow{2}{*}{Graph analytics \cite{beamer2017gapbenchmarksuite}} 
& Vertex / edge properties 
& 1 GB -- $10^2$ TB 
& 512 B -- 8 KB 
& Years 
& Yes 
& Latency \\ \cline{2-7}

& Intermediate states 
& 1 GB -- $10^1$ TB
& 512 B -- 4 KB 
& Minutes 
& No 
& Latency \\ \hline

\multirow{3}{*}{GNN training \cite{shao2024distributed,zheng2022bytegnn}} 
& Graph structure 
& $10^1$ GB -- $10^1$ TB 
& 512 B -- 8 KB 
& Years 
& Yes 
& Latency \\ \cline{2-7}

& Feature vectors 
& $10^2$ GB -- $10^2$ TB  
& 512 B -- 8 KB 
& Years 
& Yes 
& Both \\ \cline{2-7}

& Intermediate activations 
& $10^1$ GB -- $10^2$ TB
& 512 B -- 8 KB 
& Minutes 
& No 
& Throughput \\ \hline

\multirow{3}{*}{LLM training \cite{liu2024deepseek, achiam2023gpt}} 
& Input corpus 
& $10^3$ GB -- $10^1$ PB 
& 1 MB -- $10^1$ GB 
& Years 
& Yes 
& Throughput \\ \cline{2-7}

& Model weights 
& $10^1$ GB -- 1 TB 
& 1 MB -- $10^1$ GB 
& Years 
& Yes 
& Both \\ \cline{2-7}

& Optimizer states 
& $10^2$ GB -- $10^2$ TB 
& 1 MB -- $10^1$ GB 
& Days--Months 
& Yes 
& Throughput \\ \hline

LLM inference \cite{qin2025mooncake, liu2025lmcache} 
& KV cache 
& $10^2$ TB -- $10^1$ PB 
& 8 KB -- 4 MB 
& Years 
& Yes 
& Latency \\ \hline

\end{tabular}
}
\vspace{0pt}
\caption{Storage requirements of representative GPU workloads.}
\vspace{-15pt}
\label{tab:workload-storage}
\end{table*}

Our main \textbf{contributions} can be summarized as follows:

\begin{itemize}[leftmargin=10pt]
\setlength{\itemsep}{5pt}


\item We deeply review existing GPU-AFA systems and reveal that the CPU-centric design can significantly compromise I/O performance. To address this, we build \emph{GNStor}, the first end-to-end GPU-AFA system design that supports GPU-native AFA access while completely removing CPU from the remote I/O path for extremely high performance.

\item We design \emph{GNoR}, a GPU-centric NoR software stack that enables scalable remote SSD access from GPU by employing atomic-operation-based synchronization and efficient memory allocator to maximize GPU parallelism.


\item We propose \emph{deEngine}, which decentralizes and offloads AFA I/O-critical functionalities (e.g., access control and metadata maintenance) into SSD firmware, facilitating efficient sharing of AFA volumes at low cost.

\end{itemize}

%% file: sections/background.tex
\subsection{Storage Requirement of GPU Application}
\label{sec:bg_gpustorage}
GPU has gradually replaced CPU as the primary computing platform for a wide range of emerging applications, such as graph analytics \cite{nguyen2013lightweight,wang2017gunrock,pan2017multi}, recommender systems \cite{resnick1997recommender,burke2011recommender,wang2022merlin}, and large language model (LLM) \cite{achiam2023gpt,liu2024deepseek,team2023gemini}.
These applications are typically data-intensive, imposing stringent requirements on the storage system (cf. Table \ref{tab:workload-storage}).
%
We summarize primary storage requirements as follows:
(1) \emph{Large capacity}: 
These applications operate on massive datasets. For example, graph neural network (GNN) training \cite{shao2024distributed,zheng2022bytegnn} requires high-capacity storage to accommodate hundreds of TBs of adjacency matrices, feature vectors, and intermediate data. LLM training pipelines \cite{liu2024deepseek, achiam2023gpt} also rely on a large corpus that can reach PB scale;
(2) \emph{Efficient sharing}: Data is typically shared across multiple GPU clients. For instance, corpus and model parameters of LLMs are shared among multiple GPUs when performing distributed training. In LLM inference, KV cache \cite{qin2025mooncake, liu2025lmcache} is reused across multiple requests from different GPU instances to reduce computational overhead;
(3) \emph{Strong durability}: Certain data demands strong durability guarantees. In particular, long-running training jobs periodically generate large checkpoints, including model parameters and optimizer states, that must be reliably persisted in the long term to prevent costly recomputation in case of failures;
(4) \emph{High performance}: These datasets are frequently accessed by GPU during computation, making I/O performance critical. For example, training processes repeatedly load the corpus for computation, while inference workloads demand low-latency access to shared data (e.g., KV cache).

\subsection{All-Flash Array and SSD Architecture}
\label{sec:bg_afassd}
\begin{figure}
     \centering
     \includegraphics[width=1\linewidth]{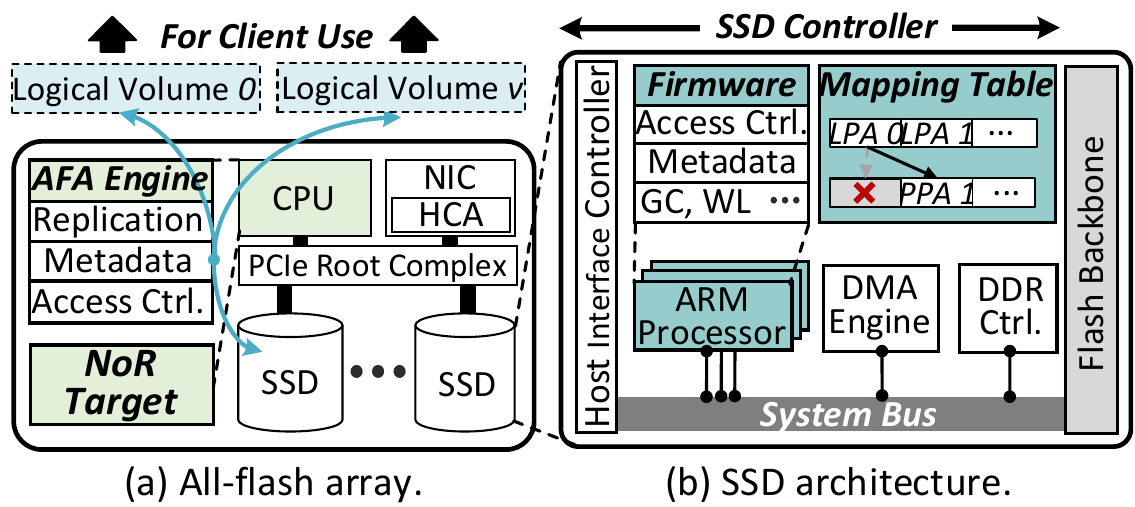}
    \vspace{-15pt}
    \caption{Architectures of AFA and SSD.\label{fig:background}}
    \vspace{-10pt}
\end{figure}

\noindent \textbf{All-flash array.} 
Due to the stringent storage requirements (i.e., large capacity, efficient sharing, strong durability, and high performance), GPU applications increasingly resort to dedicated all-flash array (AFA) \cite{pirahandeh2015energy,curry2010lightweight,khasymski2012use,scalaafa}. 
AFA is a type of storage form that bundles multiple SSDs into an array to aggregate their capacity and throughput (cf. Figure \ref{fig:background}\textcolor{fcolor}{a}). 
\emph{AFA engine} is the core of AFA software stack, which takes charge of several critical functionalities.
Specifically, it abstracts the storage space of underlying SSDs into \emph{logical volumes} for client using and enables data sharing through access control to each volume.
AFA engine also enforces data durability through redundancy mechanisms (e.g., replication), and guarantees crash consistency by properly orchestrating and persisting metadata (e.g., mapping table between addresses of logical volumes and SSDs).

\noindent \textbf{SSD architecture.} 
Figure \ref{fig:background}\textcolor{fcolor}{b} illustrates a typical architecture of modern SSDs \cite{PM1743}, which consists of an SSD controller running SSD firmware, as well as a flash backbone composed of multiple NAND flash dies.
NAND flash only supports out-of-place updates due to its physical attributes \cite{micheloni2010inside}. 
To hide this complexity from users, SSD firmware internally constructs an indirection layer, known as \emph{flash translation layer (FTL)}, to remap incoming write requests (i.e., \emph{logical page address, LPA}) to new flash pages (i.e., \emph{physical page address, PPA}) and invalidate the stale pages.
SSD firmware persists this mapping information with special hardware \cite{wu2022understanding,zhou2021remap}, enabling it to be recovered after system corruption or reboot.
Moreover, SSD firmware checks incoming I/O requests to prevent errors (e.g., access illegal address) \cite{ahn2019key,lee2017fessd}. 

\subsection{Storage Access Path of GPU-AFA System}
\label{sec:bg_iopath}
\noindent \textbf{CPU-centric access.} 
Despite the essential role of GPU in computation, existing GPU–AFA systems follow a CPU-centric I/O path.
On the GPU node, CPU is responsible for submitting I/O requests to AFA with a certain remote storage protocol.
It retrieves data from AFA to local host memory through NIC, and then forwards it to GPU memory for computation (or vice versa).
Even with optimizations such as NVIDIA GPUDirect \cite{gpudirect}, which enables direct data transfers between GPUs and PCIe devices (e.g., NICs), CPU remains critical for orchestrating I/O requests \cite{bam}.
On the AFA node, AFA engine also runs on the CPU, managing logical volumes, forcing data redundancy, and maintaining metadata. All I/O requests must pass through this CPU-based software stack.
Between the GPU and AFA nodes, \emph{NVMe over RDMA (NoR)} protocol is commonly used for high-performance access.
In a standard NoR path, the \emph{initiator} (e.g., CPU in GPU node) packs NVMe commands into NoR capsules and sends them over the RDMA to the \emph{target} (e.g., CPU in AFA node). The target then parses these capsules, translates them into block I/O operations, and issues NVMe commands to the storage devices. 
Such CPU-mediated designs suffer from a detour in I/O path and incur substantial CPU-GPU interaction overhead, becoming the performance bottleneck of high-concurrency GPU applications \cite{bam,gofs}.

\noindent \textbf{CPU bypass techniques.}
To address the inefficiency of CPU-centric approaches, great efforts have been paid to explore CPU bypass techniques.
On the GPU node, GPU-centric local storage stacks are developed. 
At the file system layer, GoFS \cite{gofs} and GeminiFS \cite{qiu2025geminifs} facilitate GPU applications to directly issue POSIX-like file system requests. At the device driver layer, libnvm \cite{libnvm} and BaM \cite{bam} expose the doorbell registers \cite{nvmebase} of NVMe SSDs into the GPU address space, enabling GPUs to directly submit I/O requests to local SSDs via PCIe memory-mapped I/O (MMIO) \cite{papagiannis2020optimizing}.
On the AFA node, modern NICs typically support NoR target offloading technique, which delegates NoR capsule handling to the host channel adapter (HCA) hardware \cite{sun2025scalio}. HCA can directly interact with the SSD via PCIe peer-to-peer (P2P) communication, avoiding CPU intervention.

\subsection{Challenges}
\label{sec:bg_challenge}
Despite the effectiveness of these CPU-bypass techniques, adopting them into GPU-AFA systems for end-to-end I/O acceleration remains challenging.
We summarize the prominent obstacles as follows:

\noindent $\bullet$ \emph{Lack of GPU-centric remote storage software stack \textbf{(Challenge 1)}.}
Existing GPU-centric storage solutions \cite{libnvm,bam} primarily target local storage (i.e., NVMe over PCIe), where the software stack is relatively simple without the involvement of network communication. 
In contrast, enabling remote storage access requires introducing a high-performance network layer, especially RDMA in NoR, which significantly complicates I/O stacks. 
For example, RDMA requires the registration of memory regions (MR) to authorize NIC to directly access the physical memory of I/O buffers. MR registration involves pinning memory pages and notifying NIC with system calls, which incurs long latency \cite{wang2025finemem}. 
Therefore, efficient RDMA MR management is critical in the NoR-based I/O path. 
Moreover, the execution models of CPU and GPU are fundamentally different. GPUs adopt a hierarchical threading model (e.g., warps, thread blocks, and grids) with thousands of concurrent threads, while existing NoR implementations are designed for CPU (typically enclosing tens of threads) with heavyweight lock-based synchronization.
Simply porting the existing CPU-oriented design to GPU would lead to severe contention on shared resources (e.g., queue pairs), resulting in excessive synchronization overheads and poor scalability 
(e.g., 86.6\% throughput degradation in our test).

\noindent $\bullet$ \emph{Lack of decentralized AFA engine in CPU-bypass system \textbf{(Challenge 2)}.}
In a fully CPU-bypassed architecture, we expect GPU-initiated AFA I/O can be delivered directly to each SSD via NIC's HCA, sidestepping the conventional CPU-resident AFA engine. 
While this design choice succeeds in shortening I/O path, simply removing the AFA engine from the storage software stack means the ignorance of the critical AFA functionalities, such as access control and metadata maintenance.
As a result, concurrent AFA accesses from multiple GPU clients can easily compromise system correctness (e.g., conflicting writes on the same storage space).
A straightforward remedy is shifting all these responsibilities to GPU clients. However, this design requires each GPU client to frequently coordinate with others for access control.
GPU client also needs to persist metadata through sophisticated mechanisms (e.g., log and checkpoint \cite{gray1981recovery}).
These extra operations impose significant synchronization and management overhead on GPU, limiting the performance of GPU-AFA systems (e.g., 72.0\% throughput loss in our evaluation).

%% file: sections/overview.tex
Tackling the aforementioned challenges, we propose \emph{GNStor}, a GPU-native AFA system that enables GPU to directly access remote AFA without CPU intervention in the I/O critical path, thereby maximizing end-to-end performance. 

\begin{figure}
    \centering
    \includegraphics[width=1\linewidth]{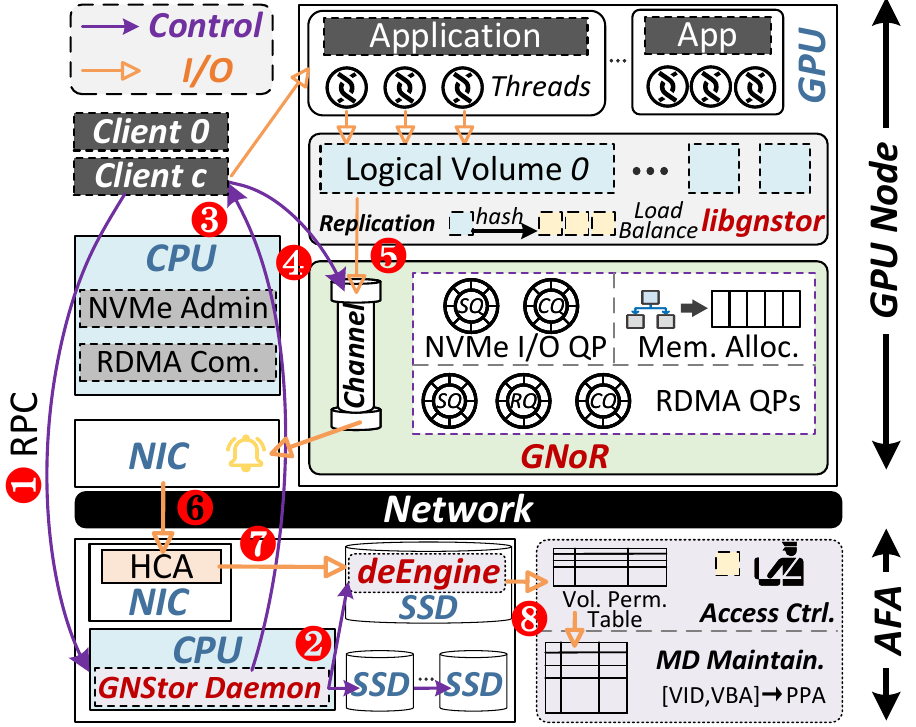}
    \vspace{-18pt}
    \caption{Overview of GNStor.\label{fig:overview}}
    \vspace{-10pt}
\end{figure}

\noindent \textbf{Base components.}
Figure \ref{fig:overview} illustrates the architecture of GNStor, which contains three key components: \emph{GNoR}, \emph{deEngine}, and \emph{GNStor daemon}.
GNoR is a high-performance GPU-centric NoR software stack that facilitates \emph{point-to-point} remote SSD access directly from GPU (solution of \emph{\textbf{Challenge 1}}).
GNoR is built from the ground up to exploit the massive parallelism of GPU architecture. 
Specifically, GNoR remains non-I/O-critical tasks and data structures (e.g., keep-alive handling and NVMe admin queues) on the CPU, as they are rarely called and have a minor impact on steady-state performance.
In contrast, GNoR migrates I/O-critical components (e.g., NVMe I/O queues and RDMA queues) to the GPU and encapsulates them into \emph{channels}, which serve as the basic I/O concurrency abstraction for GPU kernels.
Within each channel, GNoR replaces the conventional lock-based synchronization with fine-grained atomic operations to enable parallel I/O at low cost. 
In addition, GNoR minimizes RDMA MR registration overhead through pre-registered memory pools and efficient memory management.
To facilitate multiple GPU clients to access multiple AFA volumes (i.e., \emph{many-to-many}) correctly without invoking CPU in the I/O critical path as the centralized coordinator, GNStor proposes a decentralized AFA engine design (solution of \emph{\textbf{Challenge 2}}).
Infrequent AFA management tasks  (e.g., volume initialization and teardown) remain on the AFA CPU as \emph{GNStor daemon}, while I/O-critical operations are decomposed and offloaded to the firmware of each SSD as \emph{deEngine}. 
Our key insight is that many AFA I/O-critical tasks (e.g., access control and metadata persistence) overlap with functionalities already implemented in SSD firmware (cf. $\S$ \ref{sec:design_deengine}). Based on this observation, deEngine chooses to offload these tasks to each SSD by extending the SSD firmware, enabling correctness without introducing additional cost.
For better usability, GNStor exposes an expressive library, named \texttt{libgnstor}, for GPU programs, supporting synchronous and asynchronous I/O as well as batched operations.

\noindent \textbf{Workflows.}
When a GPU client requires storage, it first requests the GNStor daemon via CPU proxy (e.g., RPC \cite{bloomer1992power}) to create a new \emph{logical volume} (\redcircled{1}). 
Client can also request access to an existing logical volume for data sharing.
After validating the client’s identity, GNStor daemon propagates the access control information of the volume to the deEngines on all underlying SSDs (\redcircled{2}) and then acknowledges the client (\redcircled{3}).
Subsequently, client initializes GNoR channels to establish parallel I/O paths to every remote SSD (\redcircled{4}).
Afterward, client can issue I/O requests to the volume with \texttt{libgnstor}, which employs a hash-based load-balance strategy to determine which SSDs will handle the I/O requests (\redcircled{5}).
Each SSD I/O request is packed as a NoR command capsule and transmitted through the GNoR channel to the AFA NIC (\redcircled{6}).
Upon arrival, the NIC’s HCA parses the NoR capsule into NVMe command and forwards it to SSD with PCIe P2P communication (\redcircled{7}).
Meanwhile, HCA performs data transfer between GPU memory and SSD with RDMA one-sided operations.
Upon reaching the SSD, the NVMe command is processed by the deEngine, which validates the legitimacy of commands, executes corresponding I/O, and updates metadata (i.e., mapping tables, \redcircled{8}).

%% file: sections/design.tex
\subsection{GNStor Daemon: Volume Management}
\label{sec:design_daemon}
GNStor manages the storage space of multiple SSDs and exposes logical volumes for client to use.
Volume management is handled by the \emph{GNStor daemon}, which runs on the CPU of the AFA node. 
It can also be placed on dedicated manager servers for distributed storage system.
When a GPU client requires new storage space, it issues a request to the GNStor daemon via RPC \cite{bloomer1992power} to create a new logical volume. 
It can also specify a replica factor (i.e., the number of replicas, 2 by default) for data redundancy.
Upon receiving the request, the daemon first verifies the client’s identity and then allocates a unique volume identifier (VID) and a hash factor (cf. $\S$ \ref{sec:design_deengine}) for the volume.
The daemon then propagates the volume metadata (including the VID, hash factor, client ID, volume capacity, and replication factor) to all underlying SSDs by issuing customized \texttt{VOLUME ADD} commands. Each SSD stores the metadata in both its local DRAM and flash backbone (cf. $\S$ \ref{sec:bg_afassd}) as \emph{volume permission table}, which is later used by deEngine for access control during I/O processing. 
Afterward, the daemon returns the volume metadata to the client. 
For subsequent I/O operations, the client piggybacks the VID and client ID in each NoR command, allowing SSDs to validate access permissions (cf. $\S$ \ref{sec:design_deengine}).  
%
Client requesting access to an existing volume for data sharing has similar procedure (i.e., call GNStor daemon who informs underlying SSDs with a customized \texttt{VOLUME CHMOD} command and then returns the corresponding volume metadata to the client).
Note that, to ensure consistency, at most one client is allowed to hold write permission to a volume at any time. In our design, write permission is managed using a lease-based mechanism \cite{trach2020t}, where clients must periodically (e.g., 5 minutes) apply write permissions from the daemon.  
When a client releases a volume or relinquishes access permission, it also notifies the daemon via RPC. The daemon then updates volume permission table on SSDs with customized \texttt{VOLUME DELETE/CHMOD} command and then acknowledges the client.
GNStor assumes all clients in the GPU-AFA system are trusted (e.g., in private clusters). Access control is primarily designed to support data sharing with consistency (i.e., accessing correct SSD location that stored the demanded data and avoiding different clients occupying the same storage space), rather than defending against malicious attacks. 
In untrusted environments (e.g., public cloud), GNStor can be extended by associating cryptographic keys with each volume and requiring key-based authentication during I/O processing \cite{gofs}.

\subsection{GNoR: GPU-Centric NoR I/O Path}
\label{sec:design_gnor}
The goal of GNoR is to pave an expressway for GPU-centric remote storage access. 
GNoR re-architects the NoR software stack with consideration of GPU programming paradigms.
Specifically, GNoR introduces \emph{channels} as the basic abstraction for I/O concurrency and replaces heavyweight locks with lightweight atomic operations to enable scalable inter-thread synchronization. 
In addition, GNoR uses pre-registered memory pools and a GPU-friendly allocator to reduce the overhead of RDMA MR registration.

\begin{figure}
    \centering
    \includegraphics[width=1\linewidth]{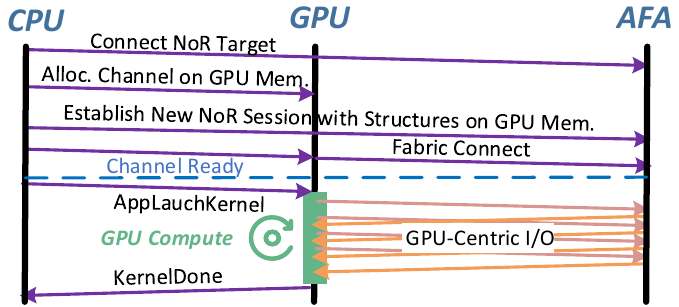}
    \vspace{-15pt}
    \caption{Initialization and I/O procedures of channel.\label{fig:channel}}
    \vspace{-10pt}
\end{figure}

\noindent \textbf{Channel.}
GNoR adopts a GPU-centric, CPU-assisted principle to partition functionalities required by the NoR specification \cite{nvmebase}. 
Specifically, CPU is responsible for tasks that are not on the I/O-critical path, such as establishing RDMA connections with remote targets and handling keep-alive messages. 
In contrast, all I/O-critical functionality are moved to the GPU.  
To be specific, GNoR bundles all necessary data structures required for issuing and completing NoR I/O requests into a unified structure named \emph{channel}, which serves as the basic abstraction for parallel I/O handling.
A channel encloses an NVMe I/O queue pair, RDMA queue pairs, addresses of NIC doorbell registers, pre-registered memory buffers, and some auxiliary state (e.g., queue tail pointer). These data structures reside in GPU memory to facilitate efficient access by GPU.
As shown in Figure \ref{fig:channel}, during system initialization, CPU establishes the NoR connection with the remote target on AFA and creates the NVMe admin queue pair. 
Afterward, GNoR can construct channels. 
For each channel, GNoR allocates channel data structures on GPU memory and starts a new NoR session with queues residing on GPU memory.
Subsequently, the pointer of channel is transferred to the GPU, which takes over the channel.  
GPU then completes the remaining setup, including pre-posting \emph{RDMA receive requests (Recvs)} and issuing the required \texttt{Fabrics Connect} command \cite{nvmebase}. 
Afterward, channel is ready and GPU can directly issue I/O requests through the channel. 
Specifically, GPU encapsulates I/O requests as NoR command capsules, submits them to the RDMA send queue, and then rings the NIC doorbell register via PCIe MMIO write to trigger network transmission.
Afterward, GPU periodically polls the RDMA completion queue to wait NVMe \emph{completion entries (CQEs)} which will arrive on the buffers pointed by RDMA Recvs. 
Finally, GPU parses the NVMe CQEs, re-posts RDMA Recvs, and triggers the corresponding asynchronous callbacks if needed.

\noindent \textbf{GPU-friendly memory allocator.}
RDMA one-sided operations require memory region (MR) registration to authorize the NIC to access the physical memory space of I/O buffers, which is a costly operation \cite{wang2025finemem}. 
To remove this overhead from the I/O critical path, GNoR adopts a pre-allocation and pre-registration strategy. Specifically, each channel maintains a pre-registered memory pool in GPU memory, from which I/O buffers are allocated dynamically during execution. 
However, efficiently managing this memory pool under massive GPU concurrency is non-trivial. 
A straightforward approach is to inherit the buddy allocator \cite{subramani2002selective} used in existing CPU-oriented NoR stacks \cite{cpulibnvmf}. However, it relies on complex data structures (e.g., hierarchical free lists) and lock-based synchronization, making it unsuitable for high-concurrency GPU scenarios.
Another promising approach is to use a bitmap-based fixed-size allocator, where each slot in the bitmap represents a fixed-size memory block (e.g., 4 KB). GPU threads acquire slots via atomic operations (i.e., compare and swap, CAS \cite{cas}) and use the corresponding memory blocks. While simple and lock-free, this approach suffers from fragmentation over time. As a result, large I/O requests may not find contiguous memory blocks and must be served by multiple disjoint buffers, leading to multiple RDMA operations for data transfer per I/O request, increasing end-to-end latency.
Our key insight is that different GPU I/O workloads exhibit relatively predictable I/O size distributions (i.e., a small set of sizes, cf. Table \ref{tab:workload-storage}).
Based on this insight, GNoR introduces a GPU-friendly multi-level allocator. 
Instead of maintaining a fully memory hierarchy to support all I/O sizes with sophisticated structures, GNoR allocator partitions the memory pool into a small number of size levels (e.g., 4 KB, 64 KB, and 1 MB), each managed by a bitmap. 
This design preserves the lock-free and low-overhead nature of bitmap allocators while significantly avoiding fragmentation.
Specifically, during allocation, requests are served from the memory level with the closest matching size to minimize internal fragmentation. The allocator also prioritizes obtaining contiguous blocks in the chosen memory level. 
Larger blocks can be selectively split to satisfy smaller allocation, while deallocation opportunistically merges adjacent free blocks to restore larger blocks. 
All these operations are implemented with fine-grained atomic primitives (e.g., CAS \cite{cas}), ensuring scalability under massive GPU concurrency.
Note that when the memory pool is insufficient and cannot satisfy allocation, it dynamically expands by allocating and registering a larger space (e.g., scales with a factor of 2 by default).

\begin{figure}
    \centering
    \includegraphics[width=0.95\linewidth]{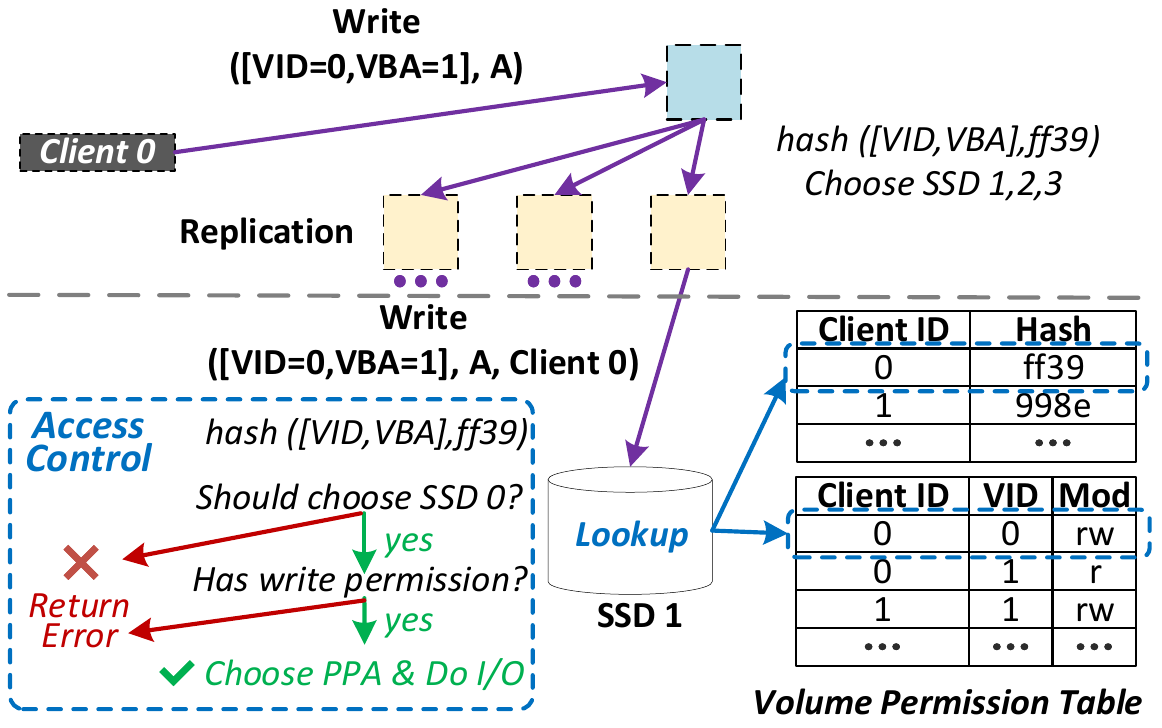}
    \vspace{-10pt}
    \caption{Access control of deEngine.\label{fig:access}}
    \vspace{-15pt}
\end{figure}

\subsection{deEngine: Decentralized AFA Engine}
\label{sec:design_deengine}
GNoR enables point-to-point direct I/O from GPU to remote SSD via NIC. Building on this foundation, deEngine supports many-to-many access of AFA without invoking CPU in the I/O critical path. 
In traditional AFA system, AFA engine runs on the CPU. All requests must traverse the CPU software stack, where the centralized AFA engine performs I/O-critical functions such as access control and metadata maintenance. 
The design prolongs the I/O path and suffers from poor scalability caused by the contention on the centralized CPU.
%
A straightforward alternative is to move these AFA functionalities to clients, similar to prior work \cite{sun2025scalio}. However, this approach introduces substantial synchronization overhead across clients for access control and requires GPUs to maintain mapping table, which becomes a performance bottleneck under high-concurrency scenarios (cf. $\S$ \ref{sec:bg_challenge}).  
Our key insight is that many I/O-critical responsibilities of AFA engine overlap with functionalities already implemented in the SSD firmware. 
Specifically, AFA engine needs to check the permission of client's I/O requests while SSD firmware must also validate the legitimacy of user's I/O requests to prevent errors (e.g., access illegal address). 
Moreover, AFA engine maintains mappings from volume-level addresses to \emph{SSD logical addresses (LPA)}, while SSD FTL maintains mappings from SSD LPA to \emph{physical locations (PPA)}. 
Based on this observation, GNStor decomposes the AFA engine and offloads I/O-critical functionalities to SSDs by integrating them into the SSD firmware (i.e., deEngine). This design completely eliminates CPU intervention in I/O path, thereby achieving higher performance.

\noindent \textbf{Access control.}
As shown in Figure \ref{fig:access}, when a client issues an I/O request to a volume, it first selects a set of target SSDs using a hash-based load-balance function \cite{karger1997consistent} over the VID and the \emph{block address in volume (VBA)}, i.e., \texttt{hash([VID,VBA],factor)}.
More sophisticated strategies (e.g., CRUSH in Ceph \cite{weil2006crush}) can also be used here to achieve load balancing in a more complex environment.
For write requests, multiple SSDs are selected according to the replication factor to ensure sufficient fault tolerance. 
%
Upon receiving a request, each deEngine independently verifies whether it is the valid target SSD by recomputing the same hash function as GPU client and checking permissions with its local volume permission table. If valid, deEngine executes the I/O operation and updates mapping table accordingly.
deEngine employs weight round robin algorithm for I/O scheduling, which is the default method in most commercial SSDs \cite{woo2021d2fq}.
Note that in GNStor I/O path, the client only needs to specify [VID,VBA] for each I/O. The final storage location (i.e., PPA) is determined by each SSD (i.e., SSDs are the coordinator). 
Therefore, GNStor prevents the error of different clients occupying the same physical storage space.
Also, clients can access the correct data location without resorting to the centralized AFA engine for table lookup (i.e., SSDs do lookup).

\begin{figure}
    \centering
    \includegraphics[width=0.85\linewidth]{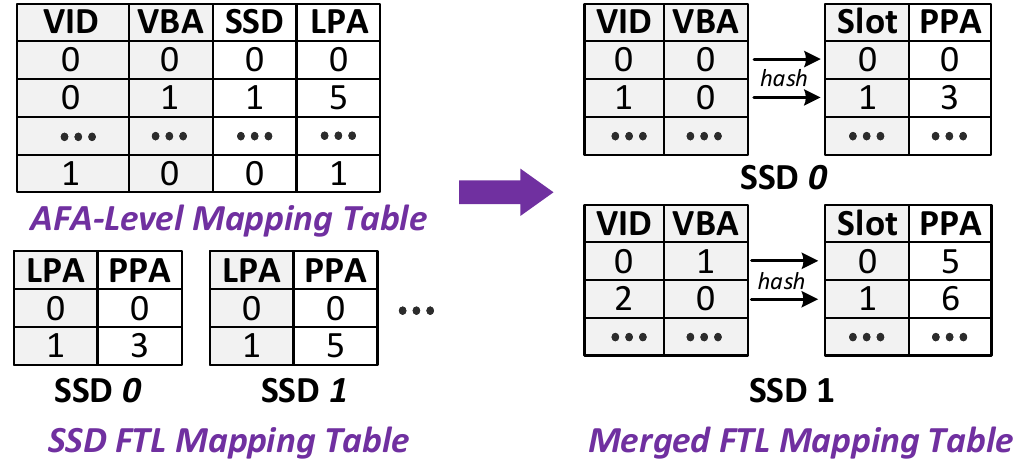}
    \vspace{-7pt}
    \caption{Merged FTL mapping table in deEngine.\label{fig:map}}
    \vspace{-15pt}
\end{figure}

\noindent \textbf{Metadata persistence and recovery.}
In conventional GPU-AFA systems, address mapping is performed in two stages: from volume addresses (i.e., [VID,VBA]) to SSD number and SSD logical addresses (LPA), and then from SSD LPA to physical locations (PPA). In GNStor, these two mappings are unified within the SSD FTL (cf. Figure \ref{fig:map}). Specifically, FTL mapping table is modified from a conventional LPA-to-PPA to a [VID,VBA]-to-PPA mapping.  
By merging these mappings to each deEngine on SSDs, GNStor eliminates the need for maintaining AFA-level mapping tables on the GPU clients. 
Note that VID and VBA do not need to be stored. GNStor uses the cuckoo hash \cite{pagh2004cuckoo} to find a mapping table slot for accommodating the PPA of [VID,VBA].

Modern SSDs provide built-in power loss protection (PLP) mechanisms to ensure the durability of FTL metadata. 
For example, to tolerate sudden power failures, enterprise SSDs are equipped with power hold-up circuits and capacitors \cite{samsungplp}, which allow buffered metadata in SSD DRAM to be flushed to persistent flash backbone upon power loss. As a result, critical metadata can be reliably recovered after system reboot.  
%
In GNStor, since AFA-level mapping tables have been integrated into the SSD-resident FTL, they naturally benefit from SSD-internal persistence and recovery mechanisms. This design eliminates the need for additional logging or checkpointing in the AFA layer.  
When a GPU client recovers or migrates, it can re-establish its identity with the GNStor daemon. The daemon then returns the corresponding volume metadata, after which the client reconstructs GNoR channels and resumes I/O operations.  
When the AFA system reboots, each SSD first restores its local mapping tables via its built-in metadata recovery process. It then recovers the volume permission table which is persisted in flash backbone (cf. $\S$ \ref{sec:design_daemon}),
Afterward, GNStor daemon reconstructs global state by retrieving volume permission table from SSDs.
Subsequently, clients can reconnect and resume I/O following the aforementioned workflow (i.e., communicating with daemon and then reconstructing channels).  
%
When SSD fails, GNStor relies on extra replicas on other SSDs to recover both data and metadata. Note that the volume permission table is replicated across all SSDs. Therefore, GNStor maintains sufficient fault tolerance to preserve its reliability. With the volume permission table, GNStor can determine which data needs to be migrated to the spare SSD (i.e., performing hash function to know which data is stored on the damaged SSD).

\begin{figure}
    \centering
    \includegraphics[width=1\linewidth]{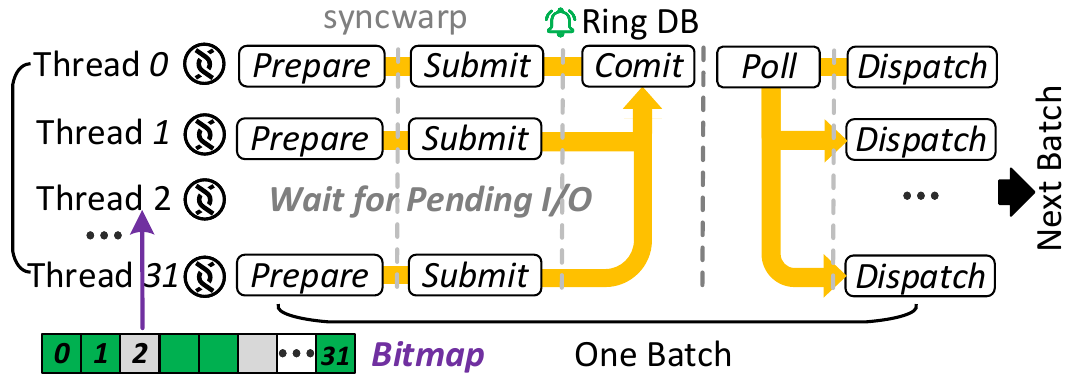}
    \vspace{-18pt}
    \caption{Illustration of batched I/O.\label{fig:batch}}
    \vspace{-12pt}
\end{figure}

\subsection{Batched I/O and libgnstor}
\noindent \textbf{Batched I/O.}
GNStor supports batched I/O to fully exploit the parallelism of GPU architecture. 
As shown in Figure \ref{fig:batch}, within a thread block, GNoR maintains a bitmap in shared memory (i.e., \texttt{\_\_shared\_\_} in CUDA) to track the status of batched requests. The bitmap contains multiple slots (typically a multiple of 32 to align with GPU warp size) and assigns one slot to each thread.  
These threads are grouped in warps (i.e., 32 threads per group).
During submission, threads in a warp get I/O buffer from the allocator (cf. $\S$ \ref{sec:design_gnor}) and prepare I/O requests in parallel.
They then use atomic operations (i.e., CAS \cite{cas}) to append their command capsules to the tail of the RDMA send queue. 
Subsequently, each thread sets its corresponding slot in the bitmap to indicate the pending command. 
After all requests are submitted, a designated thread of the warp (e.g., \texttt{thread 0}) rings the NIC doorbell with MMIO \cite{papagiannis2020optimizing} to trigger RDMA transmission.  
For completion, a designated thread (e.g., \texttt{thread 0}) polls the RDMA completion queue and dispatches NVMe CQE (in RDMA Recv buffer) to the corresponding threads, which handle results, execute callbacks, and clear the associated bitmap slots. 
These threads are synchronized with warp-level primitives (e.g., \texttt{syncwarp}) after each stage (cf. Figure \ref{fig:batch}).
Note that after a poll, some I/O requests may not have completed, so the corresponding slots will not be cleared. In the next batch, the associated threads should not submit new command (e.g., \texttt{thread 2}).
This batched execution model aligns with the single-instruction-multiple-thread (SIMT) nature of GPUs (i.e., try to have different threads in a warp execute the same instructions in parallel), reduces warp divergence, and enables efficient concurrency among threads, thereby maximizing I/O throughput.

\begin{figure}
    \centering
    \includegraphics[width=1\linewidth]{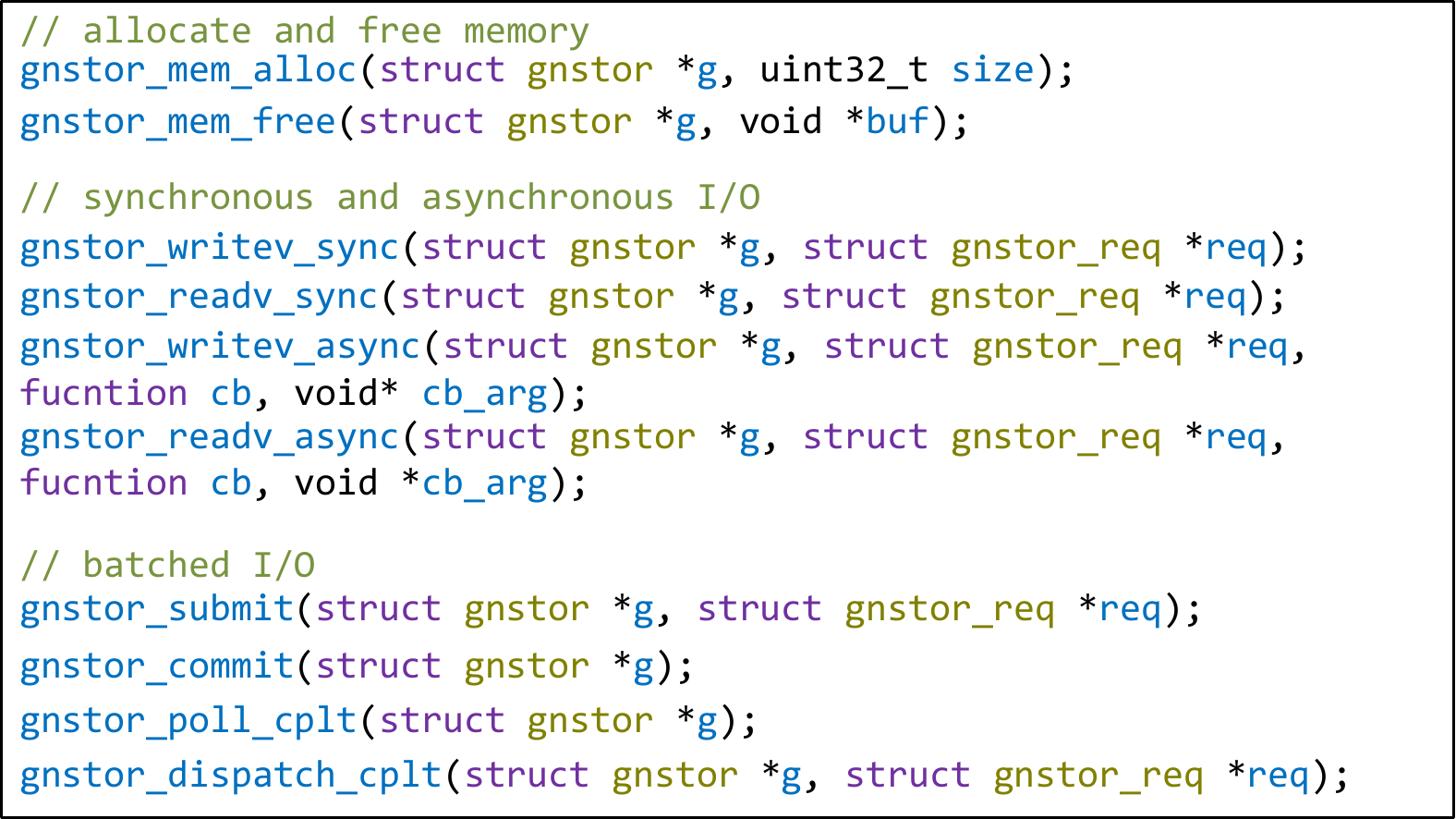}
    \vspace{-18pt}
    \caption{APIs delivered by \texttt{libgnstor}.\label{fig:code}}
    \vspace{-12pt}
\end{figure}

\noindent \textbf{libgnstor.}
For better practicality, GNStor is packaged as a library (named \texttt{libgnstor}) and provides a rich set of APIs for GPU kernels. Figure \ref{fig:code} illustrates parts of these APIs, including memory allocation/free interfaces, as well as synchronous and asynchronous I/O operations. 
GNStor supports asynchronous I/O requests by maintaining a callback function table in GPU memory. 
Once the request is completed, a designated thread will look up the table and trigger the callback.
It should be noted that batched I/O is not provided as a standalone function. This is because, in CUDA programming, \texttt{\_\_device\_\_} function can only be called by a single thread instead of assigning an entire warp. Therefore, \texttt{libgnstor} provides \texttt{submit}, \texttt{commit}, \texttt{poll}, and \texttt{dispatch} interfaces, respectively, corresponding to the stages in Figure \ref{fig:batch}.

\subsection{Implementation Details}
GNoR is implemented with 9 K LOC based on NVIDIA DOCA 3.3.0 \cite{DOCA} and CUDA 12.9 \cite{CUDA} software framework, which requires no hardware modification.
We support AFA management functions (i.e., GNStor daemon) with 3 K LOC in SPDK 25.09 \cite{spdk}.
All of our modifications to NVMe protocols are based on NVMe Base Specification 2.3 \cite{nvmebase} and NVMe Command Set Specification 1.2 \cite{NVMeCommandSet}. The customized \texttt{VOLUME ADD/DELETE/CHMOD} commands (cf. $\S$ \ref{sec:design_daemon}) are implemented as NVMe admin commands. 
VID and client ID reuse the leftmost 32 bits (16 bits for each) of the start logical block address (SLBA) of NVMe I/O commands. 
This design supports up to 16,384 clients, each with 16,384 volumes, and each volume can hold up to 16 TB (i.e., $2^{32}$ $\times$ 4 KB).
The modification to SSD firmware (i.e., deEngine) is implemented in a popular SSD emulator \cite{kim2023nvmevirt} with 2 K LOC.

%% file: sections/evaluation.tex
\subsection{Experimental Setup}
\label{sec:exp_setup}
\noindent \textbf{Methodology.}
We evaluate GNStor designs on a platform consisting of a GPU node and an AFA node, which is connected via Mellanox ConnectX-7 NICs \cite{cx7nic}, delivering 200 Gbps throughput (21.6 GB/s goodput in RDMA). 
The GPU node is equipped with a 96-core AMD EPYC 9654 CPU \cite{amd9654}, 768 GB DDR5 DRAM, and an NVIDIA A100 GPU \cite{a100gpu} with 40 GB device memory. 
The AFA node contains a 26-core Intel Xeon 5320 CPU \cite{IntelXeon5320} and 256 GB DDR4 DRAM. 
We use NVMeVirt, an accurate SSD emulator, to evaluate our deEngine design on SSD firmware. 
We configure emulated SSDs as high-performance storage devices, each delivering  6.8 GB/s and 4.8 GB/s peak read and write bandwidth, respectively. These settings match with commercial SSD products \cite{samsung980pro}.
We get configurations of the deEngine from Xilinx Kintex UltraScale+ KU15P FPGA with 300 MHz clock \cite{xilinxfpga}.
It takes 276 ns more time to calculate the hash function for address mapping and access control (cf. $\S$ \ref{sec:design_deengine}). We cross-validate the performance model used in our emulator with this result.
The key configurations are listed in Table \ref{tab:config}.

\begin{table}
\centering
\renewcommand{\arraystretch}{1}
\resizebox{\linewidth}{!}{%
\setlength{\tabcolsep}{6pt}
\begin{tabular}{|cccc|cc|}
\hline
\multicolumn{2}{|c|}{\textbf{GPU node}} &
  \multicolumn{2}{c|}{\textbf{AFA node}} &
  \multicolumn{2}{c|}{\textbf{Software}} \\ \hline
\multicolumn{1}{|c|}{\multirow{2}{*}{\textbf{CPU}}} &
  \multicolumn{1}{c|}{AMD EPYC 9654} &
  \multicolumn{1}{c|}{\multirow{2}{*}{\textbf{CPU}}} &
  Intel Xeon 5320 &
  \multicolumn{1}{c|}{\textbf{Ubuntu}} &
  22.04 \\ \cline{2-2} \cline{4-6} 
\multicolumn{1}{|c|}{} &
  \multicolumn{1}{c|}{1 x 96 core / 2.4 GHz} &
  \multicolumn{1}{c|}{} &
  1 x26 core / 2.2 GHz &
  \multicolumn{1}{c|}{\textbf{Linux}} &
  5.4.0 \\ \hline
\multicolumn{1}{|c|}{\textbf{Mem.}} &
  \multicolumn{1}{c|}{768 GB DDR5} &
  \multicolumn{1}{c|}{\multirow{3}{*}{\textbf{SSDs}}} &
  4 KB R/W: 3250/2980 MB/s &
  \multicolumn{1}{c|}{\textbf{CUDA}} &
  12.9 \\ \cline{1-2} \cline{4-6} 
\multicolumn{1}{|c|}{\multirow{2}{*}{\textbf{GPU}}} &
  \multicolumn{1}{c|}{NVIDIA A100} &
  \multicolumn{1}{c|}{} &
  64 KB R/W: 6988/4950 MB/s &
  \multicolumn{1}{c|}{\multirow{2}{*}{\textbf{\begin{tabular}[c]{@{}c@{}}GPU\\ driver\end{tabular}}}} &
  \multirow{2}{*}{575.57.08} \\ \cline{2-2} \cline{4-4}
\multicolumn{1}{|c|}{} &
  \multicolumn{1}{c|}{40 GB device memory} &
  \multicolumn{1}{c|}{} &
  4 SSDs / 2 replicas &
  \multicolumn{1}{c|}{} &
   \\ \hline
\multicolumn{1}{|c|}{\multirow{2}{*}{\textbf{NICs}}} &
  \multicolumn{3}{c|}{\multirow{2}{*}{\begin{tabular}[c]{@{}c@{}}Mellanox ConnectX-7 200 GbE \\ 21.6 GB/s (=22118 MB/s) goodput in RoCE \end{tabular}}} &
  \multicolumn{1}{c|}{\textbf{DOCA}} &
  3.3.0 \\ \cline{5-6} 
\multicolumn{1}{|c|}{} &
  \multicolumn{3}{c|}{} &
  \multicolumn{1}{c|}{\textbf{SPDK}} &
  25.09 \\ \hline
\end{tabular}
}
\vspace{0pt}
\caption{System configurations.\label{tab:config}}
\vspace{-22pt}
\end{table}

\noindent \textbf{Baselines.}
We compare GNStor with two other designs.
(1) \texttt{Basic}: conventional CPU-centric GPU-AFA system design \cite{spdk,cpulibnvmf}, in which CPU takes charge of orchestrating remote I/O requests and forwarding data to GPU memory via \texttt{cudaMemcpy}. It also executes a centralized AFA engine for I/O handling in the AFA node (cf. $\S$ \ref{sec:bg_iopath});  
(2) \texttt{GD}: compared to \texttt{Basic}, it utilizes NVIDIA GPUDirect \cite{gpudirect} to enable direct data transfer between NIC and GPU memory (i.e., bypassing host memory), but still relies on CPU for I/O orchestration;  
(3) \texttt{GNStor}: a GPU-native AFA system that includes all the techniques proposed in this paper.

\begin{table}
\centering
\renewcommand{\arraystretch}{1}
\resizebox{\linewidth}{!}{%
\setlength{\tabcolsep}{4pt}
\scriptsize
\begin{tabular}{|>{\centering\arraybackslash}m{0.17\linewidth}
                |>{\centering\arraybackslash}m{0.2\linewidth}
                |>{\raggedright\arraybackslash}m{0.52\linewidth}|}
\hline
\textbf{Application} & \textbf{Workloads} & \centering\arraybackslash\textbf{Description} \\ \hline

\multirow[c]{2}{*}{\makecell[c]{\textbf{Tensor}\\\textbf{computing}}}
& Vector addition
& Compute the sum of two vectors, each containing 1 billion elements. \\ \cline{2-3}

& Matrix multiplication
& Compute the product of two $16384 \times 16384$ matrices. \\ \hline

\textbf{Data pre-processing}
& Image resize
& Resize images in ImageNet-100 \cite{imagenet100} using the bilinear interpolation algorithm. \\ \hline

\multirow[c]{3}{*}{\makecell[c]{\textbf{Graph}\\\textbf{analytics}}}
& Breadth-first search
& Conduct breadth-first search on the GAP dataset \cite{beamer2017gapbenchmarksuite}. \\ \cline{2-3}

& Connected components
& Conduct connected-components analysis on the GAP dataset \cite{beamer2017gapbenchmarksuite}. \\ \cline{2-3}

& Single-source shortest path
& Conduct single-source shortest-path analysis on the GAP dataset \cite{beamer2017gapbenchmarksuite}. \\ \hline

\textbf{LLM training}
& GPT-2
& Train a large language model (LLM) using GPT-2  \cite{gpt2}. \\ \hline
\end{tabular}
}
\vspace{0pt}
\caption{Applications tested in the evaluation.\label{tab:application}}
\vspace{-25pt}
\end{table}

\noindent \textbf{Workloads.}
We conduct evaluations with both microbenchmarks and real data-intensive GPU applications.
For microbenchmarks, we employ one client by default. This client equips one CPU thread and one GPU warp with one GNoR channel to handle I/O. 
We also vary the number of clients for scalability tests in $\S$ \ref{sec:exp_scalability}.
We use 4 SSDs with 2 replicas and examine settings with different numbers of SSDs in $\S$ \ref{sec:exp_scalability}. 
In the AFA node, for CPU-centric designs (i.e., \texttt{Basic} and \texttt{GD}), we use 8 CPU cores to run the AFA engine, aligning with the SSD/CPU-core ratio in commercial AFA products \cite{powerstore500t}.
For real-world applications, we choose tensor computing, data pre-processing, graph analytics, and LLM training (cf. Table \ref{tab:application}), as they are representative GPU workloads in production environment.

\begin{figure}
 \centering
    \includegraphics[width=1\linewidth]{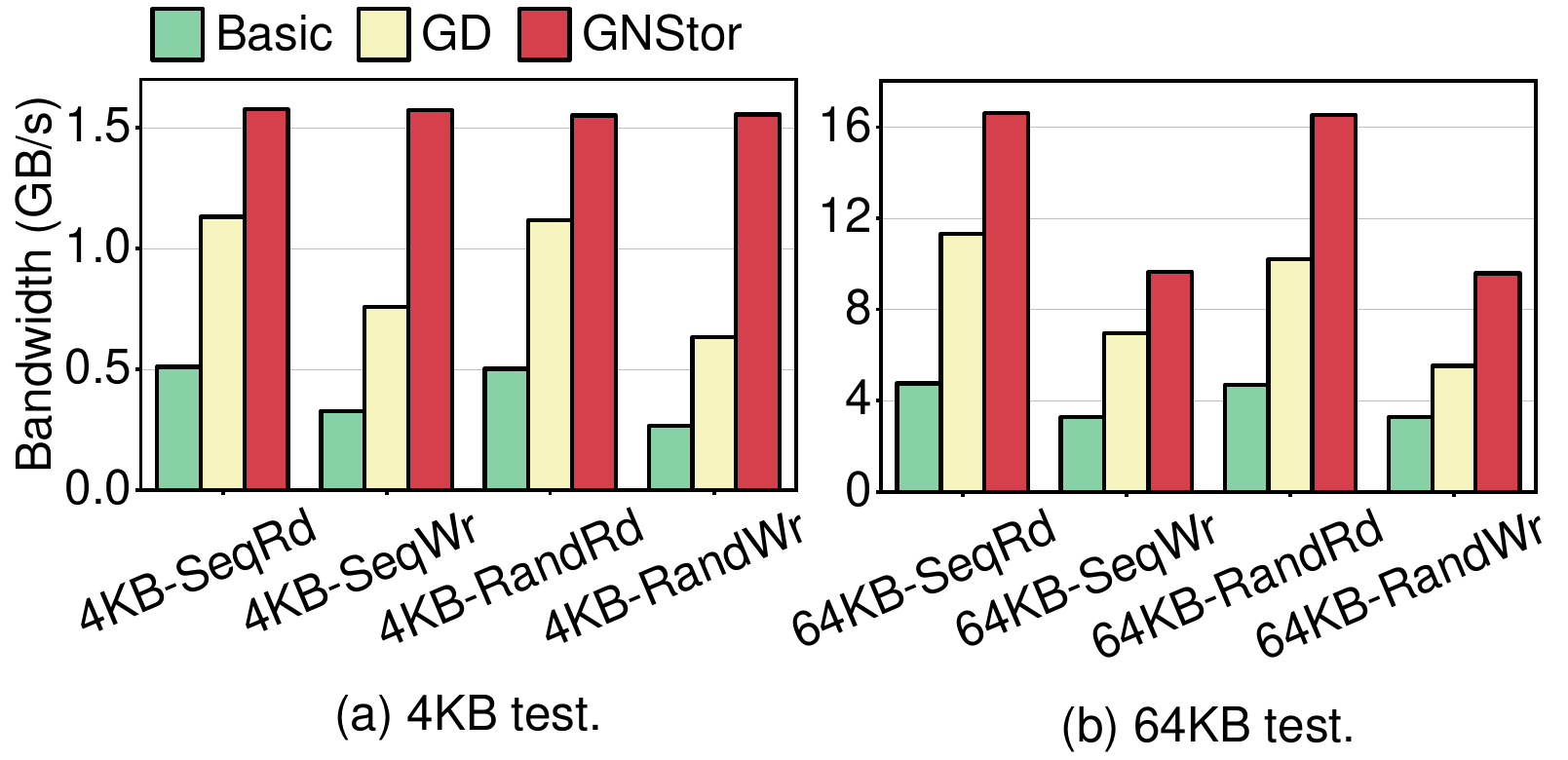}
    \vspace{-22pt}
    \caption{Throughput comparison on microbenchmarks.\label{fig:exp_micro_band}}
    \vspace{-12pt}
\end{figure}

\subsection{Overall Performance}
\label{sec:exp_overall}
\noindent \textbf{Throughput.}
Figure \ref{fig:exp_micro_band} shows the read and write throughput of different tested platforms on microbenchmarks. We set I/O depth as 32. We vary the access patterns from sequential ones to random ones and examine different I/O sizes, including 4 KB and 64 KB. \texttt{Basic} delivers only 0.5 GB/s and 0.3 GB/s read and write throughput in 4 KB tests, far from reaching the peak performance of AFA, because its CPU-centric design suffers from significant CPU-GPU interaction overhead. 
Compared with \texttt{Basic}, \texttt{GD} succeeds in improving 4 KB read and write throughput by 1.2$\times$ and 1.3$\times$. This is because \texttt{GD} enables direct data transfers between NIC and GPU memory, avoiding a detour to host memory.
\texttt{GNStor} achieves the highest throughput in all the workloads. For example, \texttt{GNStor} outperforms \texttt{Basic} and \texttt{GD} by 3.2$\times$ and 0.8$\times$ in 4 KB tests, on average. This superiority also exists in 64 KB comparisons. This is because \texttt{GNStor} not only benefits from our GPU-centric NoR software stack (i.e., GNoR) but also enjoys the decentralized deEngine design, thereby completely avoiding CPU intervention in I/O path for higher performance.

\begin{figure}
 \centering
    \includegraphics[width=1\linewidth]{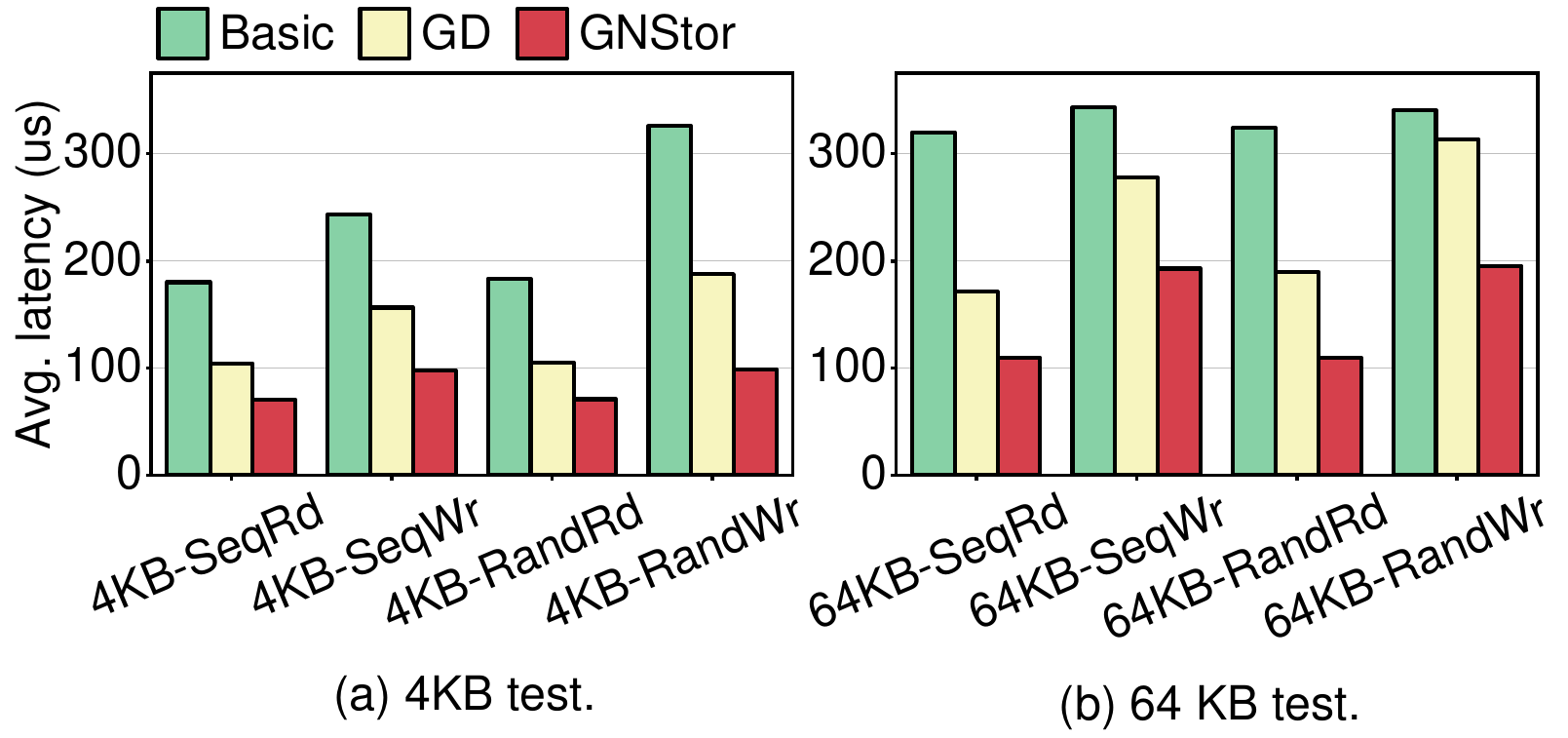}
    \vspace{-22pt}
    \caption{Latency comparison on microbenchmarks.\label{fig:exp_micro_lat}}
    \vspace{-10pt}
\end{figure}

\noindent \textbf{Latency.}
Figure \ref{fig:exp_micro_lat} illustrates the latency of different GPU-AFA systems in diverse I/O patterns.
\texttt{GD} outperforms \texttt{Basic} by 40.7\% and 26.1\% in 4 KB and 64 KB tests, respectively, echoing our findings in throughput comparison.
This is because \texttt{GD} enables data zero-copy between GPU and NIC, avoiding the time-consuming detour to host memory.
\texttt{GNStor} further reduces read and write latency by 35.7\% and 39.8\%, thanks to our fully CPU-bypass design, which paves the shortest I/O path between GPU and AFA.

\begin{figure}
 \centering
    \includegraphics[width=1\linewidth]{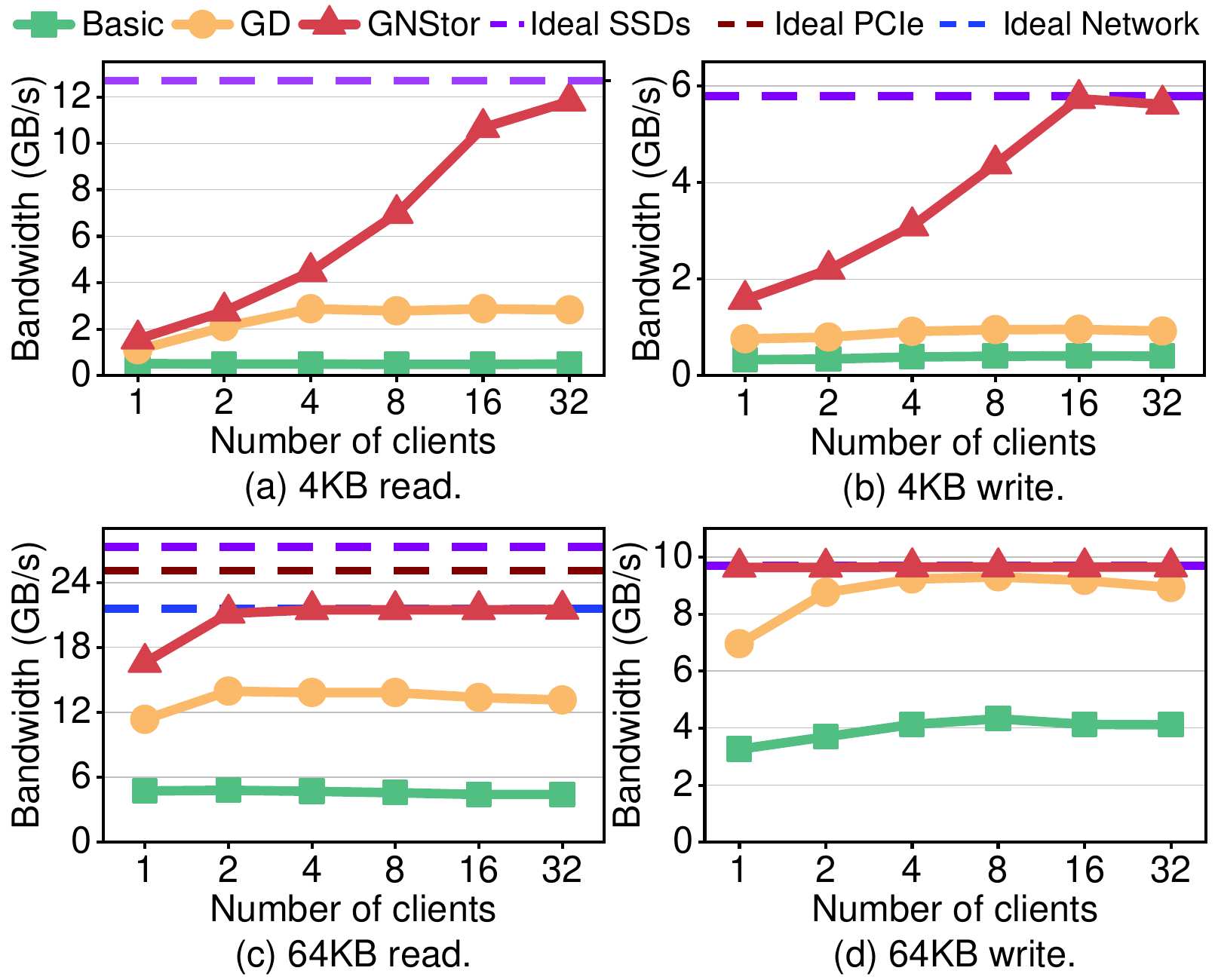}
    \vspace{-18pt}
    \caption{Scalability test with different numbers of clients.\label{fig:exp_num_clients}}
    \vspace{-5pt}
\end{figure}

\subsection{Scalability Test}
\label{sec:exp_scalability}
\noindent \textbf{Different numbers of clients.}
We further measure the scalability of different platforms by varying the number of clients from 1 to 32. For \texttt{Basic} and \texttt{GD}, we employ one CPU thread per client. For \texttt{GNStor}, we assign one GPU warp and one GNoR channel per client. 
Each client has its own logical volumes.
For reference, we also plot the ideal throughput of our testbed, including (1) the maximum goodput of network (i.e., 21.6 GB/s), which can be considered as the ideal performance of remote storage; (2) the maximum goodput of GPU's PCIe lanes (i.e., 25.1 GB/s), which can be considered as the ideal performance of local storage; and (3) the maximum throughput of SSDs. Note that we use four SSDs with two replicas in this test. Therefore, the maximum read and write throughput should be 4 $\times$ and 2 $\times$ of the peak throughput of a single SSD 
(i.e., 12.7 GB/s, 5.8 GB/s, 27.3 GB/s, and 9.7 GB/s for 4 KB read, 4 KB write, 64 KB read, and 64 KB write, respectively, cf. Table \ref{tab:config}).  
Figure \ref{fig:exp_num_clients} depicts the results. 
\texttt{Basic} gains limited benefits in multi-client scenarios, even with extra CPU threads. 
For example, it achieves only 4.4 GB/s and 4.1 GB/s read and write throughput in 64 KB tests.
This can be attributed to its CPU-centric design, in which the CPU-GPU interaction and memory copy overhead become the performance bottleneck.
\texttt{GD} is more scalable than \texttt{Basic}, thanks to its peer-to-peer data transfer in GPU node that avoids one extra copy in host memory.
However, it still relies on CPU for I/O orchestration and experiences a detour to the centralized AFA engine in storage node, which easily limits performance scalability in high-concurrency scenarios.
Therefore, \texttt{GD} only achieves 2.8 GB/s and 0.9 GB/s 4 KB read and write throughput, 22.1\% and 15.5\% of the ideal performance.
In comparison, thanks to our fully CPU-bypass design to maximize GPU parallelism, \texttt{GNStor} achieves 11.8 GB/s, 5.6 GB/s, and 9.6 GB/s throughput in 4 KB read, 4 KB write, and 64 KB write tests, nearly reaching the maximum throughput of SSDs. 
In 64 KB test, \texttt{GNStor} improves the throughput to 21.5 GB/s with only two clients, almost exhausting the network bandwidth (i.e., 99.5\%).
Moreover, it exploits 85.7\% of GPU's PCIe throughput. This result proves that our design can even rival local storage to some extent.

\begin{figure}
 \vspace{3pt}
 \centering
    \includegraphics[width=1\linewidth]{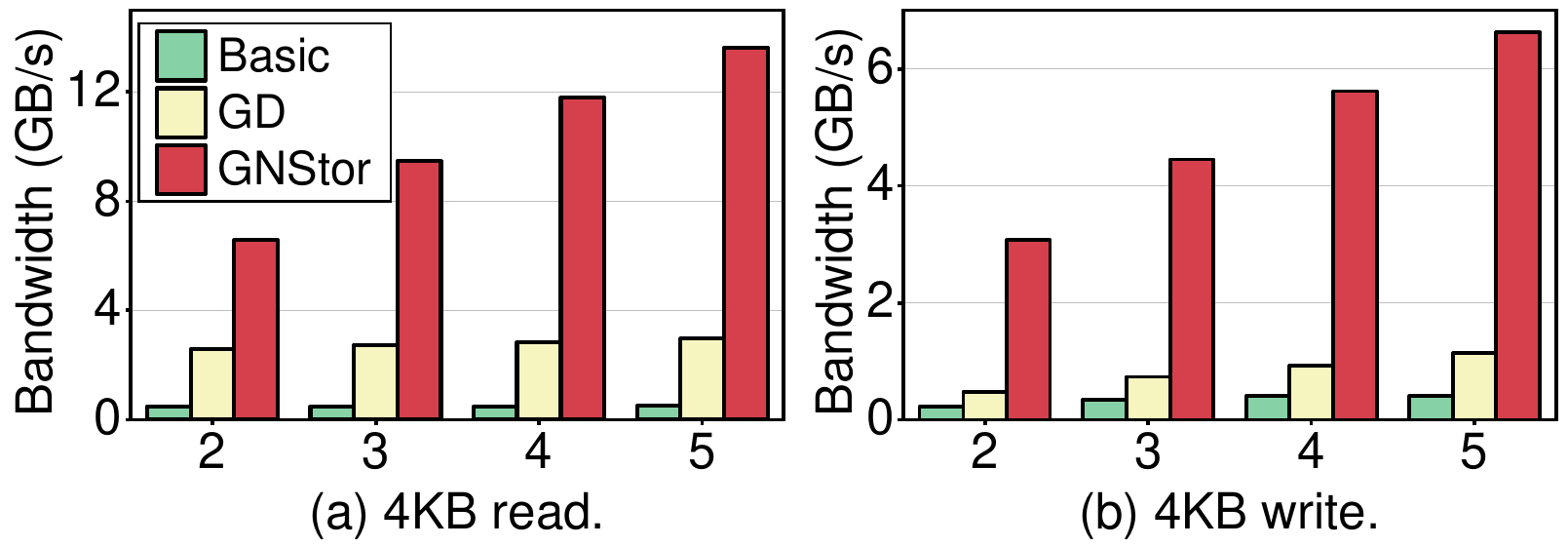}
    \vspace{-23pt}
    \caption{Scalability test with different numbers of SSDs.\label{fig:exp_num_ssds}}
    \vspace{-13pt}
\end{figure}

\begin{figure}
 \centering
    \includegraphics[width=1\linewidth]{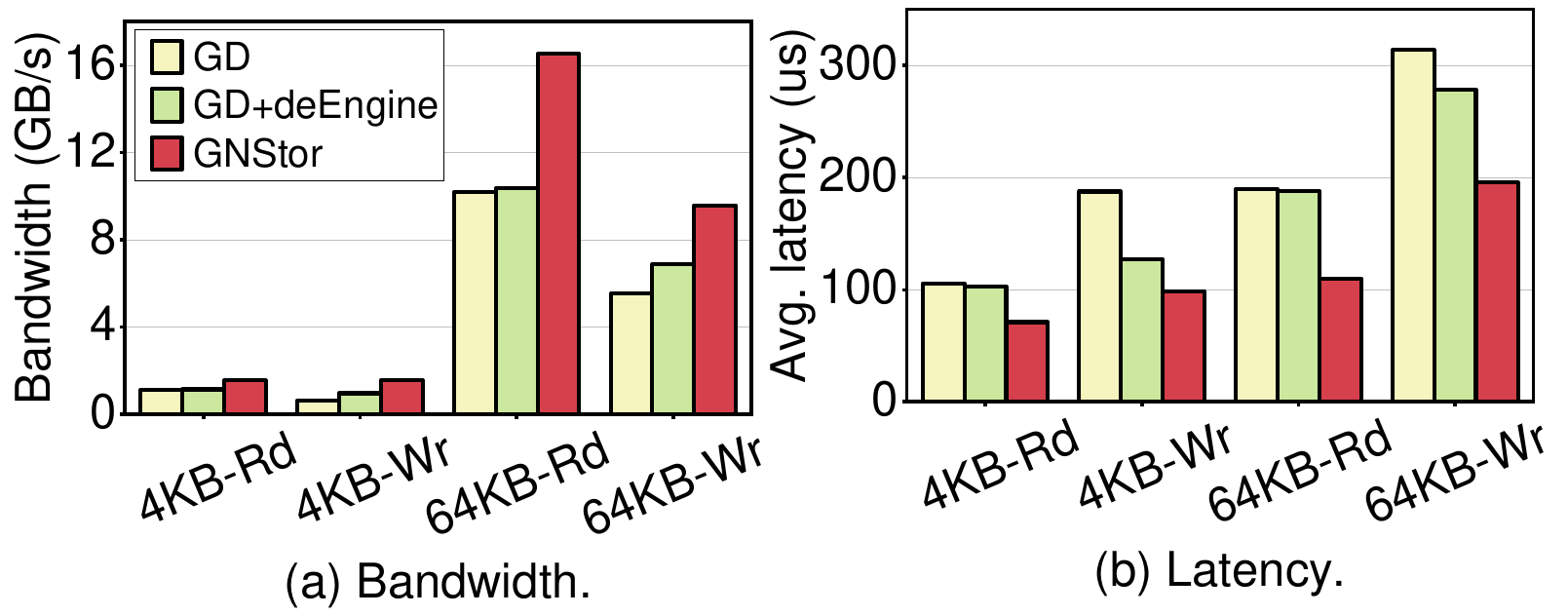}
    \vspace{-22pt}
    \caption{Ablation analysis.\label{fig:exp_ablation}}
    \vspace{-12pt}
\end{figure}

\begin{figure*}
    \centering
    \begin{minipage}[t]{0.265\textwidth} 
        \centering
        \includegraphics[width=\linewidth]{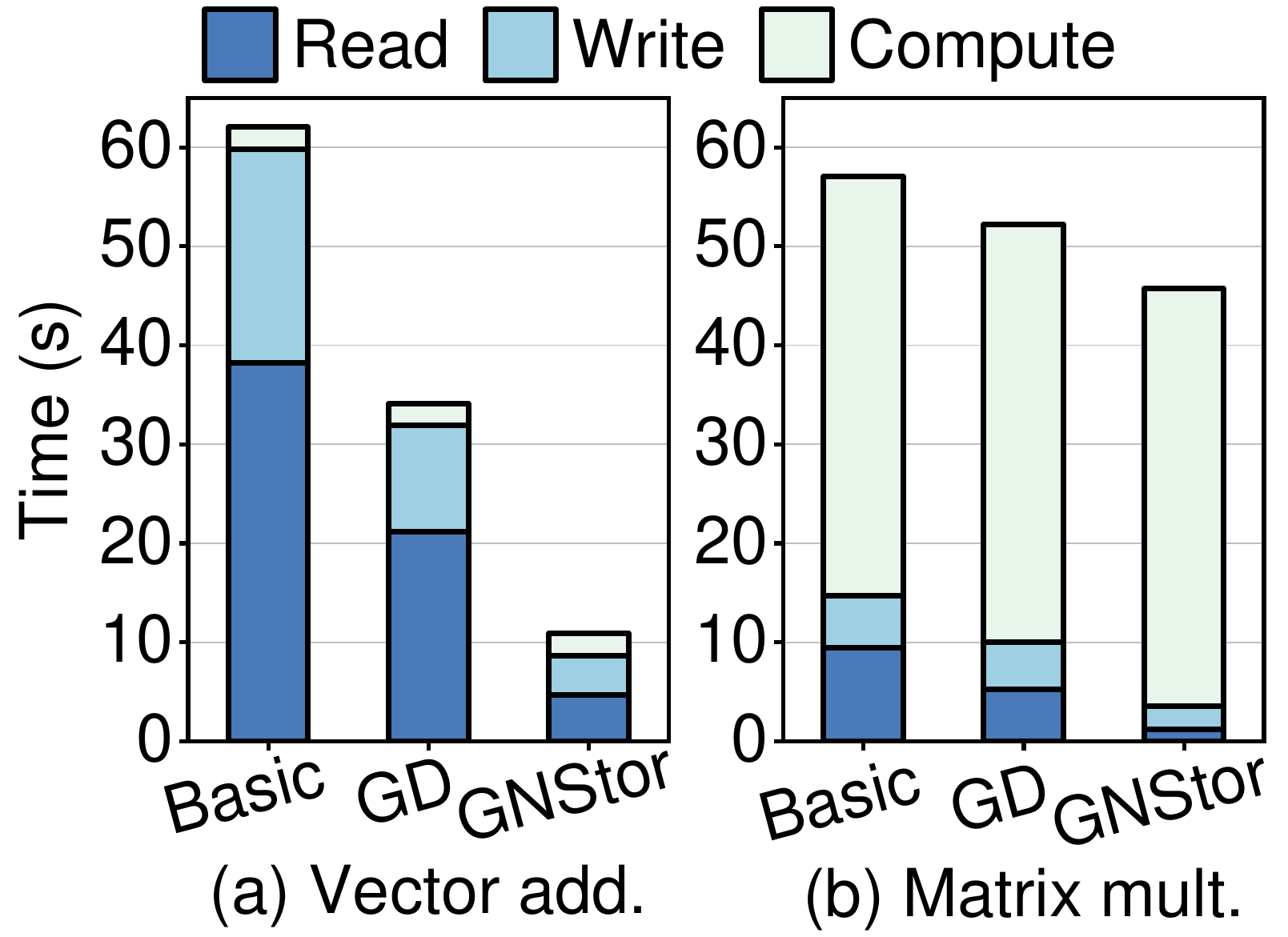}
        \vspace{-20pt}
        \captionof{figure}{Tensor computing.\label{fig:exp_app_tensorcpt}}
    \end{minipage}
    \hfill
    \begin{minipage}[t]{0.29\textwidth} 
        \centering
        \includegraphics[width=\linewidth]{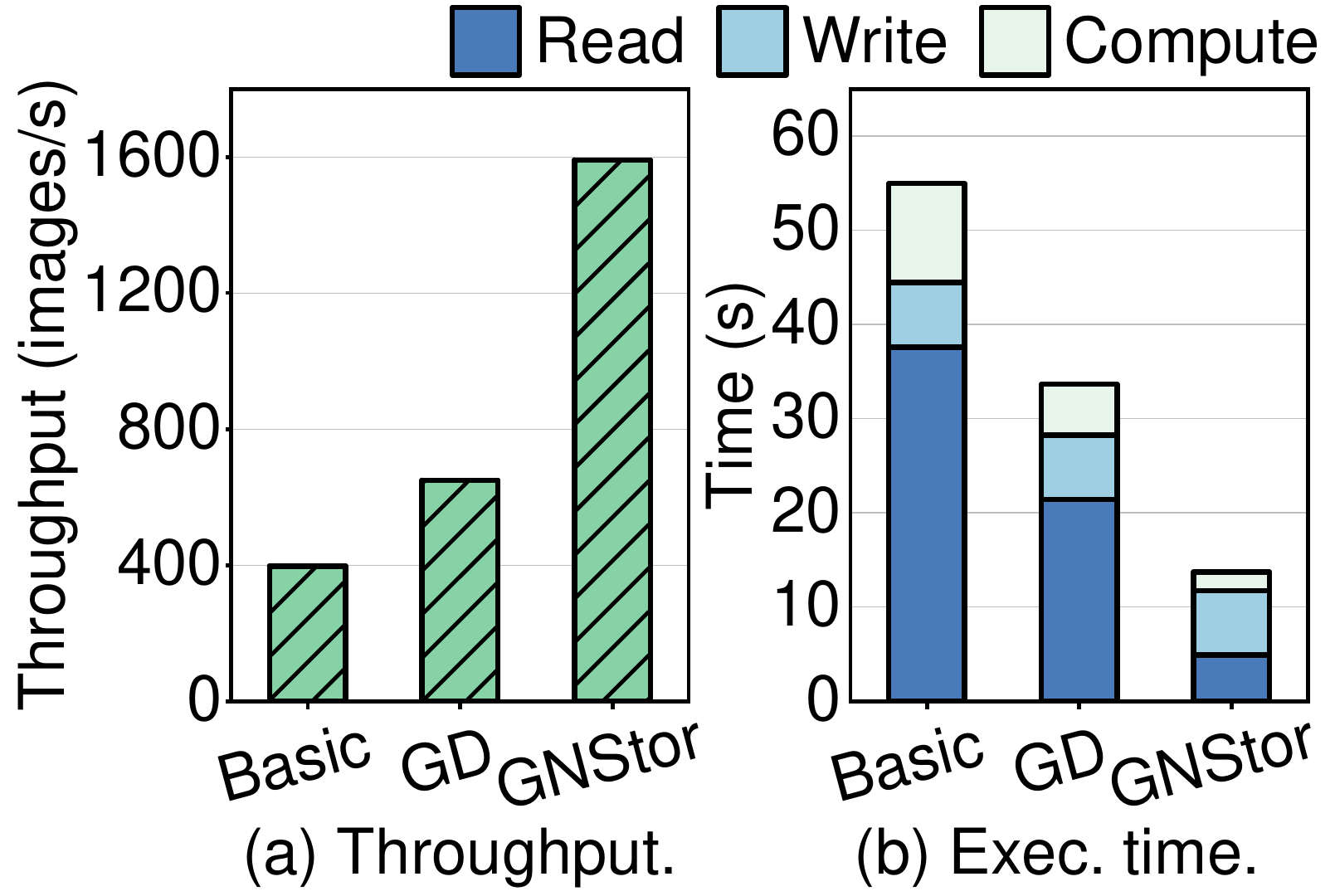}
        \vspace{-20pt}
        \captionof{figure}{Data pre-processing.\label{fig:exp_app_datapreprocess}}
    \end{minipage}
    \hfill
    \begin{minipage}[t]{0.415\textwidth} 
        \centering
        \includegraphics[width=\linewidth]{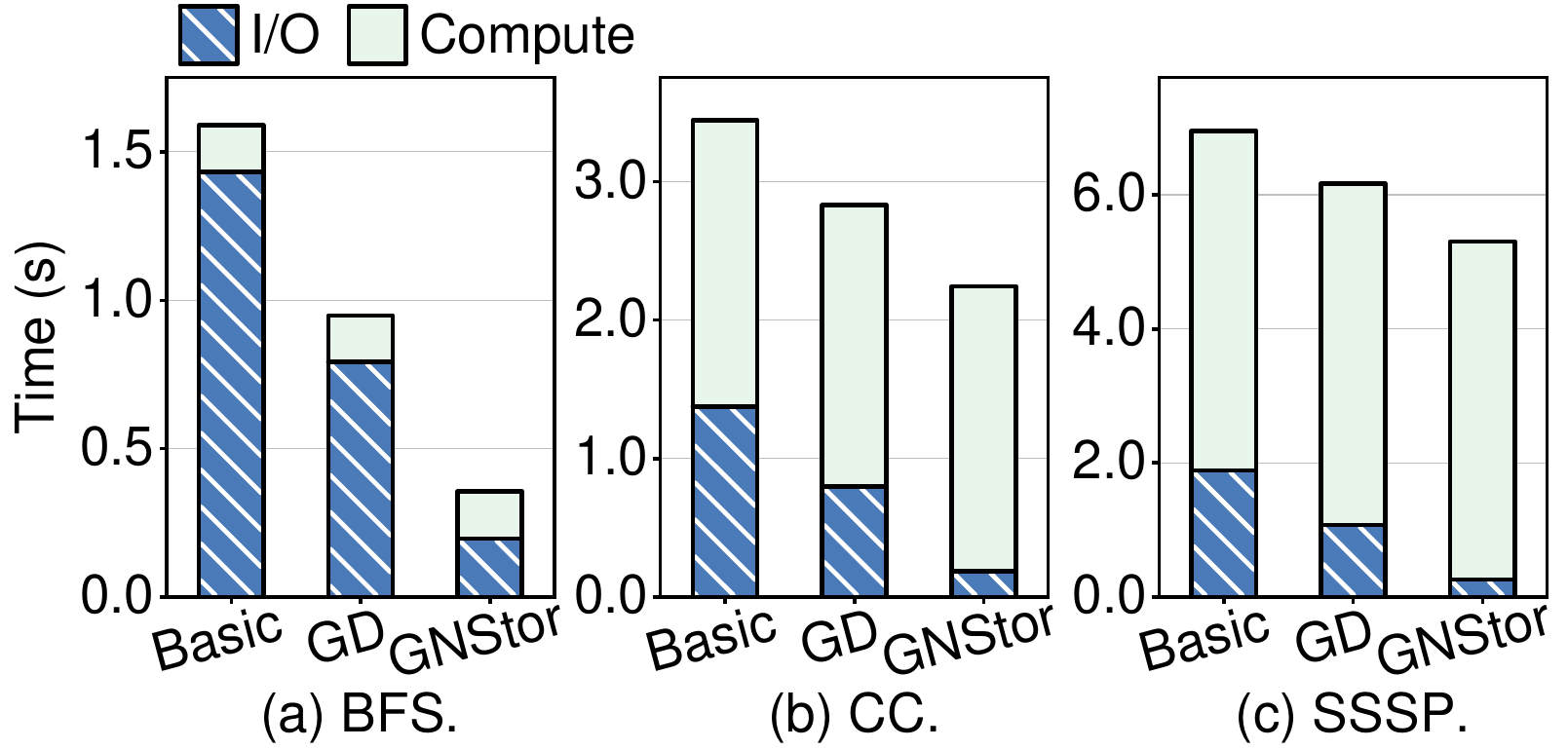}
        \vspace{-20pt}
        \captionof{figure}{Graph analytics.\label{fig:exp_app_graph}}
    \end{minipage}
    \vspace{-0pt}
\end{figure*}

\noindent \textbf{Different numbers of SSDs.}
We now evaluate our design with different numbers of SSDs. We set the number of replicas as 2 in all the tests.
As shown in Figure \ref{fig:exp_num_ssds}, \texttt{Basic} and \texttt{GD} have minor improvement with more SSDs. 
This is because their CPU-centric lock-based design causes huge synchronization overhead, thereby limiting the throughput.
In comparison, \texttt{GNStor} exhibits good scalability in all the tests. It achieves 13.6 GB/s and 6.6 GB/s throughput in 4 KB sequential read and write tests with five SSDs (3.6$\times$ and 4.8$\times$ higher than \texttt{GD}).
We attribute this performance superiority to our GPU-friendly atomic-operation-based I/O parallelism mechanism and decentralized deEngine design.

\subsection{Ablation Analysis}
\label{sec:exp_ablation}
%
To examine how much GPU-AFA systems can benefit from each of our core designs (mainly deEngine and GNoR, cf. $\S$ \ref{sec:design}), we set an ablation experiment. We equip \texttt{GD} with our deEngine design (i.e., \texttt{GD+deEngine}) as the intermediate counterpart. 
Figure \ref{fig:exp_ablation} shows the throughput and latency comparison on different random I/O patterns.
Compared with \texttt{GD}, \texttt{GD+deEngine} achieves 49.9\% and 24.7\% higher write throughput in 4 KB and 64 KB tests, respectively. 
The latency advantages are 32.4\% and 11.1\%.
This is because deEngine offloads the AFA-level access control and metadata maintenance to SSD firmware,
which frees the centralized CPU-based AFA engine. 
deEngine design has a relatively minor impact on the read test. This is because, in \texttt{GD}, read requests do not need to update AFA-level mapping tables, avoiding frequent logs for ensuring crash consistency of the metadata. 
Therefore, it rarely reaps the benefits brought by SSD-driven metadata maintenance.
Based on \texttt{GD+deEngine}, \texttt{GNStor} further improves throughput by 49.6\% and reduces latency by 31.2\% for all workloads, on average.
This can be attributed to our GNoR design, which enables GPU-centric remote SSD access, thereby mitigating CPU-GPU interaction overhead in CPU-mediated I/O path.

\subsection{Real-World Applications}
\label{sec:exp_application}
To demonstrate end-to-end performance improvement brought by our designs, we now conduct comparisons on representative GPU applications as listed in Table \ref{tab:application}.

\noindent \textbf{Tensor computing.} 
We compare the performance of different designs by conducting vector addition and matrix multiplication workloads, which are cornerstones of numerous GPU applications. For vector addition, we use two vectors, each with one billion 8-byte elements (i.e., \texttt{double}), as input. These vectors are stored in remote AFA and should be loaded into GPU memory in batches (1 MB per batch) for computation. Afterward, the results will be written back to AFA for persistence. For matrix multiplication, we use two 16,384 $\times$ 16,384 matrices.
Figure \ref{fig:exp_app_tensorcpt} shows the comparison of execution times. We also break down the results to examine the benefit brought by I/O acceleration.
In the vector addition test, \texttt{GNStor} shortens I/O times (i.e., \texttt{Read} and \texttt{Write}) by 85.5\% and 72.8\%, compared with \texttt{Basic} and \texttt{GD}, respectively. 
This is because \texttt{GNStor} enjoys its GPU-centric remote storage stack that avoids CPU intervention, thereby significantly reducing CPU-GPU interaction overhead. Also, \texttt{GNStor} benefits from the decentralized deEngine and batched operations, further accelerating end-to-end I/O performance.
As a result, \texttt{GNStor} achieves 5.7$\times$ and 3.1$\times$ speedup over \texttt{Basic} and \texttt{GD}, respectively.
In the matrix multiplication workload, computation (i.e., \texttt{Compute}) dominates the execution time. However, \texttt{GNStor} still outperforms \texttt{Basic} and \texttt{GD} by 24.7\% and 12.4\%.

\noindent \textbf{Data pre-processing.}
Data pre-processing is the prerequisite task for many GPU workloads (e.g., model training), involving extensive read and write on storage to process the training dataset for later use. In our test, we run the bilinear interpolation algorithm \cite{gribbon2004novel} on GPU to resize images in ImageNet-100 dataset \cite{imagenet100}. These images are processed in batches (16 GB per batch).
Figure \ref{fig:exp_app_datapreprocess} shows the throughput and execution time breakdown of a batch. 
\texttt{GNStor} achieves the highest throughput (3.0$\times$ and 1.5$\times$ higher than \texttt{Basic} and \texttt{GD}) in this evaluation. This is because it can exploit the massive parallelism in GPU architecture for I/O handling and avoid the GPU-CPU interaction overhead in CPU-centric designs. As evidence, Figure \ref{fig:exp_app_datapreprocess}\textcolor{fcolor}{b} shows that \texttt{GNStor} successfully reduces 73.5\% and 58.4\% I/O time than \texttt{Basic} and \texttt{GD}.

\noindent \textbf{Graph analytics.}
We examine three representative graph analytic algorithms: breadth-first search (\texttt{BFS}), connected components (\texttt{CC}), and single-source shortest path (\texttt{SSSP}). All these workloads contain multiple iterations. In each iteration, GPU processes all vertices in the current iteration simultaneously and decides which vertices need to be used in the next iteration (e.g., adjacent ones). Then, it fetches the new vertices for the next iteration of computation.
Figure \ref{fig:exp_app_graph} presents execution time of different GPU-AFA platforms.
Thanks to our GPU-native design, \texttt{GNStor} achieves 7.3$\times$ and 4.0$\times$ I/O speedup over \texttt{Basic} and \texttt{GD} in \texttt{BFS}. This I/O acceleration converts into 77.6\% and 62.5\% shorter execution time.
\texttt{CC} and \texttt{SSSP} are more compute-intensive. 
However, \texttt{GNStor} still outperforms \texttt{Basic} and \texttt{GD} by 29.2\% and 17.3\%, on average, because of its high end-to-end storage performance.

\begin{figure}
 \centering
    \includegraphics[width=1\linewidth]{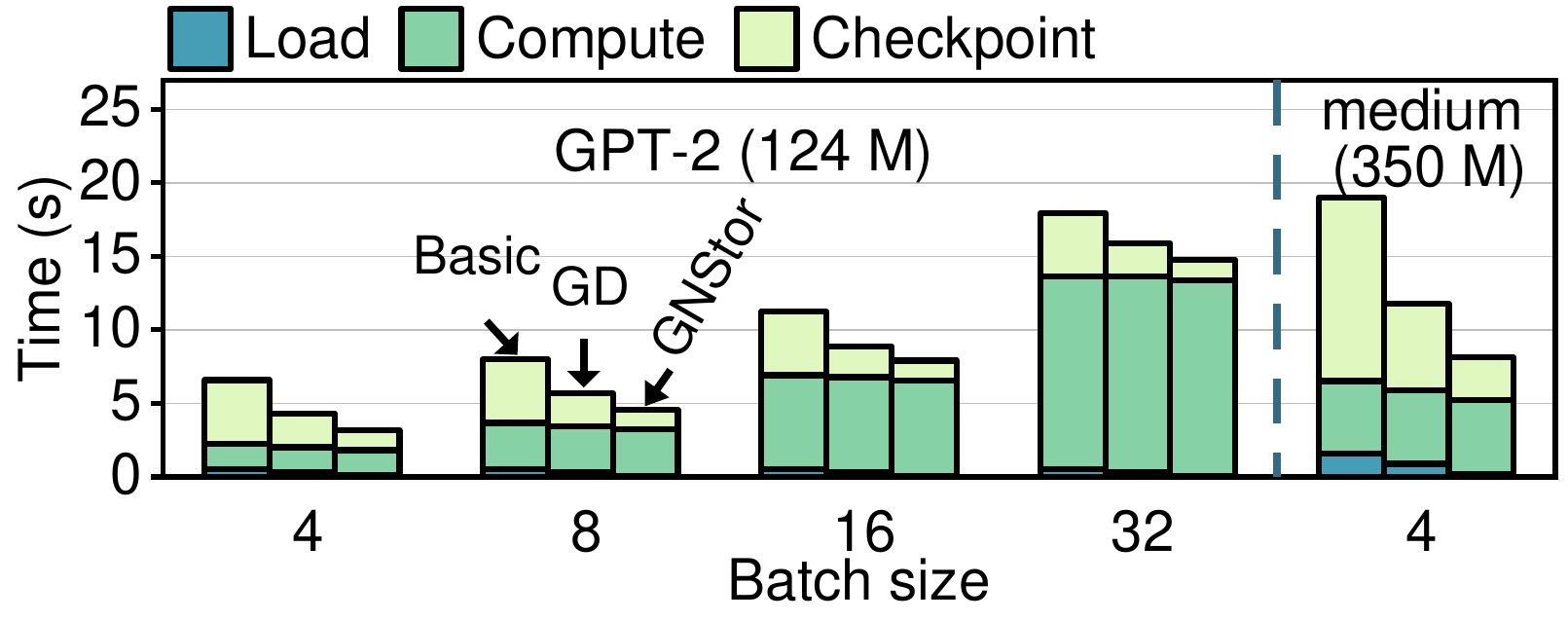}
    \vspace{-20pt}
    \caption{GPT-2 training.\label{fig:exp_app_llm}}
    \vspace{-10pt}
\end{figure}

\noindent \textbf{LLM training.}
We now evaluate GNStor with LLM training workload using GPT-2 model \cite{gpt2}. 
We vary the batch size from 4 to 32 and examine different model sizes. 
In each run, GPU loads model weights (\texttt{Load}) from remote AFA, trains 50 steps (\texttt{Compute}), and then writes back the checkpoint (including intermediate model weights and optimizer states, \texttt{Checkpoint}). Training datasets are cached in local host memory for fast loading.
As shown in Figure \ref{fig:exp_app_llm}, \texttt{GNStor} reduces execution time by 46.5\% and 26.4\% than \texttt{Basic} and \texttt{GD} when the batch size is 4, echoing our findings in previous experiments. 
This should be attributed to our GPU-native CPU-bypass design that constructs an extremely fast I/O path between GPU and remote AFA. As evidence, \texttt{GNStor} shortens \texttt{Load} and \texttt{Checkpoint} by 84.0\% and 68.7\% than \texttt{Basic}.
With a larger batch size, \texttt{Compute} becomes more dominant. Nevertheless, GNStor continues to deliver the best performance thanks to its substantially high end-to-end I/O performance.
In scenarios with larger model sizes (i.e., GPT-2 medium), the checkpoint size also increases, which further amplifies the advantage of \texttt{GNStor} (e.g., 1.3$\times$ speedup over \texttt{GD}).

\subsection{GPU Memory Footprint of GNStor}
\label{sec:exp_memory}
GNStor has integrated the AFA-layer mapping table into the SSD FTL. Therefore, for each volume, GNStor only needs to maintain a few metadata (e.g., 16-bit VID and 64-bit hash factor, cf. $\S$ \ref{sec:design_daemon}) in GPU memory.
The primary GPU memory consumption stems from the channel data structures, including NVMe queue pairs, RDMA queue pairs, pre-allocated memory pools, and auxiliary states (cf. $\S$ \ref{sec:design_gnor}). The size of queue structures is related to the maximum number of concurrent requests allowed by the channel, which is set as 128 in our default configuration, occupying approximately 50 KB per channel. Moreover, we configure an 8 MB memory pool per channel (cf. $\S$ \ref{sec:design_gnor}).
According to our evaluation (cf. $\S$ \ref{sec:exp_scalability}), 32 channels (about 258 MB memory consumption) are enough to maximize throughput of tested AFA, which is minor compared to the tens of GBs of GPU memory.



%% file: sections/relatedwork.tex

\noindent \textbf{In-SSD processing.}
As modern SSDs usually equip abundant computing resources (e.g., ARM processor and on-board DRAM), much prior work \cite{liang2019cognitive,yang2023lambda,zhou2021remap,kim2025d2fs,scalaafa,wen2024eliminating} proposes to offload tasks to SSDs. 
Cognitive SSD \cite{liang2019cognitive} accelerates deep learning workloads with in-SSD resources. 
$\lambda$-I/O \cite{yang2023lambda} proposes a general software stack to allow applications to offload data-intensive functions to the computational storage devices.
ScalaAFA \cite{scalaafa} succeeds in offloading the parity computation of stripe-based AFA (e.g., RAID 5) to the SSDs. 
In contrast, GNStor focuses on mitigating the overhead of AFA-level access control and metadata maintenance.
D2FS \cite{kim2025d2fs} integrates the metadata of file system into SSD firmware to mitigate garbage collection overhead on host CPU. 
GNStor extends this idea by taking the system overheads of the GPU-AFA I/O path into consideration.

\noindent \textbf{System designs with NoR target offloading.}
Great efforts \cite{chen2023hyq,li2023rubbledb,sun2025scalio, xu2024performance} have been taken to evolve system designs with NoR target offloading techniques.
HyQ \cite{chen2023hyq} designs a novel queue architecture for I/O scheduling in NoR target offloading scenarios.
RubbleDB \cite{li2023rubbledb} proposes a distributed database that leverages NoR target offloading to accelerate data replication.
Scalio \cite{sun2025scalio} supports multiple clients to access a shared key-value store in NoR target offloading I/O path (i.e., no centralized coordinator). These clients frequently access a shared memory area with RDMA to negotiate with each other for access control and metadata maintenance.
In contrast, GNStor chooses to offload these functionalities into SSDs and integrate them with SSD firmware, thereby avoiding inter-GPU-client synchronization overhead in high-concurrency scenarios at low cost.

\noindent \textbf{Accelerating storage access for GPU.}
Multiple studies \cite{gpudirect,zhang2015nvmmu,bergman2019spin,bam,libnvm,qiu2025geminifs,gofs,unat2024landscape,song2025cam} have been proposed to optimize I/O path for GPU. 
GPUDirect \cite{gpudirect}, NVMMU \cite{zhang2015nvmmu}, and SPIN \cite{bergman2019spin} employ PCIe peer-to-peer communication to enable direct data movement between GPU and PCIe devices (e.g., SSDs).
These designs still rely on the CPU for I/O orchestration.
BaM \cite{bam} and libnvm \cite{libnvm} map the doorbell registers of NVMe SSDs into GPU memory, allowing it to access local SSD without CPU involvement.
In contrast, GNStor enables GPU-centric access to remote AFA by properly handling network layer tasks (e.g., RDMA MR registration).
GeminiFS \cite{qiu2025geminifs} and GoFS \cite{gofs} succeed in introducing file system to GPU storage stack. These designs are complementary to GNStor as they enable more flexible data organization with high-layer abstractions (e.g., file formats).
Industry is also enthusiastically exploring to evolve GPU-centric storage. 
NVIDIA SCADA project \cite{SCADA1} encourages to optimize GPU-initiated remote storage access and is actively collaborating with SSD manufacturers to design SSDs better suited for GPU use.
GNStor aligns with this trend and holds the promise.

\noindent \textbf{Practicality and compatibility.}
The main obstacle of deploying GNStor is the modification of SSD firmware (i.e., deEngine). Limited by SSD manufacturers, it is difficult to achieve this with off-the-shelf SSDs. However, we believe in its prospects, given the substantial performance benefits brought by deEngine (cf. $\S$ \ref{sec:evaluation}) and the emerging trend of in-SSD processing. 
GNoR is a software-only solution. Although it is currently developed on NVIDIA’s software framework (i.e., CUDA \cite{CUDA} and DOCA \cite{DOCA}), its design does not depend on any specific hardware architecture. 
GNoR can be ported to GPUs and NICs from other vendors as long as the GPUs follow similar programming paradigms and the NICs can expose doorbell registers to the GPU.

%% file: sections/conclusion.tex
GPUs have become the dominant computing platform for data-intensive applications, which heavily rely on remote AFAs for scalable storage. However, existing GPU–AFA systems remain CPU-centric, leading to substantial CPU-GPU interaction overhead and I/O inefficiencies.
Tackling this issue, we propose GNStor, a GPU-native AFA system that eliminates CPU involvement from the I/O critical path. 
GNStor re-architects the NVMe over RDMA storage stack with GPU-oriented design principles, enabling GPUs to directly issue I/O to SSDs of remote AFA with high scalability. 
Furthermore, GNStor decentralizes AFA functionality by seamlessly integrating access control and metadata maintenance into SSD firmware, thereby achieving efficient AFA sharing at low cost.
Our evaluation results reveal that GNStor improves I/O throughput by 3.2$\times$ and reduces application execution time by 31.1\%, compared to state-of-the-art GPU-AFA systems.

%% file: ref.bib
@String{Computing = "Computing" }

@String{Computer = "{IEEE} Computer" }

@String{Springer = "Springer-Verlag" }

@misc{PM1743,
    author = {Samsung},
    title = {Samsung PM1743},
    howpublished = {\url{https://semiconductor.samsung.com/ssd/enterprise-ssd/pm1743/}},
    year={2023}
}

@inproceedings{zhou2021remap,
  title={$\{$Remap-SSD$\}$: Safely and efficiently exploiting $\{$SSD$\}$ address remapping to eliminate duplicate writes},
  author={Zhou, You and Wu, Qiulin and Wu, Fei and Jiang, Hong and Zhou, Jian and Xie, Changsheng},
  booktitle={19th USENIX Conference on File and Storage Technologies (FAST 21)},
  pages={187--202},
  year={2021}
}

@article{lee2017fessd,
  title={FESSD: A fast encrypted ssd employing on-chip access-control memory},
  author={Lee, Junghee and Ganesh, Kalidas and Lee, Hyuk-Jun and Kim, Youngjae},
  journal={IEEE Computer Architecture Letters},
  volume={16},
  number={2},
  pages={115--118},
  year={2017},
  publisher={IEEE}
}

@article{ahn2019key,
  title={KEY-SSD: Access-control drive to protect files from ransomware attacks},
  author={Ahn, Jinwoo and Park, Donggyu and Lee, Chang-Gyu and Min, Donghyun and Lee, Junghee and Park, Sungyong and Chen, Qian and Kim, Youngjae},
  journal={arXiv preprint arXiv:1904.05012},
  year={2019}
}

@Misc{nvmebase,
author = {NVMe},
howpublished = {\url{https://nvmexpress.org/specification/nvm-express-base-specification/}},
year={2025},
title={NVM Express Base Specification 2.3}
}

@inproceedings{papagiannis2020optimizing,
  title={Optimizing memory-mapped $\{$I/O$\}$ for fast storage devices},
  author={Papagiannis, Anastasios and Xanthakis, Giorgos and Saloustros, Giorgos and Marazakis, Manolis and Bilas, Angelos},
  booktitle={2020 USENIX Annual Technical Conference (USENIX ATC 20)},
  pages={813--827},
  year={2020}
}

@article{gray1981recovery,
  title={The recovery manager of the System R database manager},
  author={Gray, Jim and McJones, Paul and Blasgen, Mike and Lindsay, Bruce and Lorie, Raymond and Price, Tom and Putzolu, Franco and Traiger, Irving},
  journal={ACM Computing Surveys (CSUR)},
  volume={13},
  number={2},
  pages={223--242},
  year={1981},
  publisher={ACM New York, NY, USA}
}

@inproceedings{subramani2002selective,
  title={Selective buddy allocation for scheduling parallel jobs on clusters},
  author={Subramani, Vijay and Kettimuthu, Rajkumar and Srinivasan, Srividya and Johnston, Jeanette and Sadayappan, P},
  booktitle={Proceedings. IEEE International Conference on Cluster Computing},
  pages={107--116},
  year={2002},
  organization={IEEE}
}

@inproceedings{cas,
  title={Lock-free linked lists using compare-and-swap},
  author={Valois, John D},
  booktitle={Proceedings of the fourteenth annual ACM symposium on Principles of distributed computing},
  pages={214--222},
  year={1995}
}

@book{bloomer1992power,
  title={Power programming with RPC},
  author={Bloomer, John},
  year={1992},
  publisher={" O'Reilly Media, Inc."}
}

@inproceedings{woo2021d2fq,
  title={$\{$D2FQ$\}$:$\{$Device-Direct$\}$ Fair Queueing for $\{$NVMe$\}$$\{$SSDs$\}$},
  author={Woo, Jiwon and Ahn, Minwoo and Lee, Gyusun and Jeong, Jinkyu},
  booktitle={19th USENIX Conference on File and Storage Technologies (FAST 21)},
  pages={403--415},
  year={2021}
}

@misc{IntelXeon5320,
    author = {Intel},
    title = {Intel® Xeon® Gold 5320 Processor},
    howpublished = {\url{https://www.intel.com/content/www/us/en/products/sku/215285/intel-xeon-gold-5320-processor-39m-cache-2-20-ghz/specifications.html}}
}

@misc{NVMeCommandSet,
    author = {NVMe},
    title = {NVM Command Set Specification},
    howpublished = {\url{https://nvmexpress.org/specification/nvm-command-set-specification/}},
    year = {2025}
}

@misc{DOCA,
    author = {NVIDIA},
    title = {DOCA Software Framework},
    howpublished = {\url{https://developer.nvidia.com/networking/doca}},
    year = {2025}
}

@misc{CUDA,
    author = {NVIDIA},
    title = {CUDA Toolkit},
    howpublished = {\url{https://developer.nvidia.com/cuda/toolkit}},
    year = {2025}
}

@misc{cx7nic,
    author = {NVIDIA},
    title = {CONNECTX-7},
    howpublished = {\url{https://www.nvidia.com/content/dam/en-zz/Solutions/networking/ethernet-adapters/connectx-7-datasheet-Final.pdf}},
    year = {2021}
}

@misc{cpulibnvmf,
    author = {Bytedance},
    title = {NVMe over Fabrics user space initiator library},
    howpublished = {\url{https://github.com/bytedance/libnvmf}},
    year = {2024}
}

@inproceedings{karger1997consistent,
  title={Consistent hashing and random trees: Distributed caching protocols for relieving hot spots on the world wide web},
  author={Karger, David and Lehman, Eric and Leighton, Tom and Panigrahy, Rina and Levine, Matthew and Lewin, Daniel},
  booktitle={Proceedings of the twenty-ninth annual ACM symposium on Theory of computing},
  pages={654--663},
  year={1997}
}

@inproceedings{wen2024eliminating,
  title={Eliminating storage management overhead of deduplication over ssd arrays through a hardware/software co-design},
  author={Wen, Yuhong and Zhao, Xiaogang and Zhou, You and Zhang, Tong and Yang, Shangjun and Xie, Changsheng and Wu, Fei},
  booktitle={Proceedings of the 29th ACM International Conference on Architectural Support for Programming Languages and Operating Systems, Volume 2},
  pages={320--335},
  year={2024}
}

@article{pagh2004cuckoo,
  title={Cuckoo hashing},
  author={Pagh, Rasmus and Rodler, Flemming Friche},
  journal={Journal of Algorithms},
  volume={51},
  number={2},
  pages={122--144},
  year={2004},
  publisher={Elsevier}
}

@inproceedings{trach2020t,
  title={T-lease: A trusted lease primitive for distributed systems},
  author={Trach, Bohdan and Faqeh, Rasha and Oleksenko, Oleksii and Ozga, Wojciech and Bhatotia, Pramod and Fetzer, Christof},
  booktitle={Proceedings of the 11th ACM Symposium on Cloud Computing},
  pages={387--400},
  year={2020}
}

@misc{beamer2017gapbenchmarksuite,
      title={The GAP Benchmark Suite}, 
      author={Scott Beamer and Krste Asanović and David Patterson},
      year={2017},
      eprint={1508.03619},
      archivePrefix={arXiv},
      primaryClass={cs.DC},
      url={https://arxiv.org/abs/1508.03619}, 
}

@inproceedings{weil2006crush,
  title={CRUSH: Controlled, scalable, decentralized placement of replicated data},
  author={Weil, Sage A and Brandt, Scott A and Miller, Ethan L and Maltzahn, Carlos},
  booktitle={Proceedings of the 2006 ACM/IEEE conference on Supercomputing},
  pages={122--es},
  year={2006}
}

@article{unat2024landscape,
  title={The landscape of gpu-centric communication},
  author={Unat, Didem and Turimbetov, Ilyas and Issa, Mohammed Kefah Taha and Sa{\u{g}}bili, Do{\u{g}}an and Vella, Flavio and De Sensi, Daniele and Ismayilov, Ismayil},
  journal={arXiv preprint arXiv:2409.09874},
  year={2024}
}

@inproceedings{song2025cam,
  title={CAM: Asynchronous GPU-Initiated, CPU-Managed SSD Management for Batching Storage Access},
  author={Song, Ziyu and Zhang, Jie and Sun, Jie and Sun, Mo and Yang, Zihan and Zhang, Zheng and Chen, Xuzheng and Wu, Fei and Tang, Huajin and Wang, Zeke},
  booktitle={2025 IEEE 41st International Conference on Data Engineering (ICDE)},
  pages={2309--2322},
  year={2025},
  organization={IEEE}
}

@inproceedings{xu2024performance,
  title={Performance characterization of smartnic nvme-over-fabrics target offloading},
  author={Xu, Jiexiong and Qiu, Yue and Chen, Yiquan and Wang, Yijing and Lin, Wenhai and Lin, Yiquan and Zhao, Shushu and Liu, Yuqi and Wang, Ying and Chen, Wenzhi},
  booktitle={Proceedings of the 17th ACM International Systems and Storage Conference},
  pages={14--24},
  year={2024}
}

@misc{a100gpu,
    author = {NVIDIA},
    title = {A100 Tensor Core GPU},
    howpublished = {\url{https://www.nvidia.com/en-us/data-center/a100/}},
}

@inproceedings{liang2019cognitive,
  title={Cognitive $\{$SSD$\}$: A deep learning engine for $\{$In-Storage$\}$ data retrieval},
  author={Liang, Shengwen and Wang, Ying and Lu, Youyou and Yang, Zhe and Li, Huawei and Li, Xiaowei},
  booktitle={2019 USENIX Annual Technical Conference (USENIX ATC 19)},
  pages={395--410},
  year={2019}
}

@inproceedings{kim2025d2fs,
  title={$\{$D2FS$\}$:$\{$Device-Driven$\}$ Filesystem Garbage Collection},
  author={Kim, Juwon and Lee, Seungjae and Oh, Joontaek and Shin, Dongkun and Won, Youjip},
  booktitle={23rd USENIX Conference on File and Storage Technologies (FAST 25)},
  pages={337--353},
  year={2025}
}

@inproceedings{chen2023hyq,
  title={Hyq: Hybrid i/o queue architecture for nvme over fabrics to enable high-performance hardware offloading},
  author={Chen, Yiquan and Chen, Jinlong and Wang, Yijing and Chen, Yi and Jin, Zhen and Xu, Jiexiong and Fang, Guoju and Lin, Wenhai and Wei, Chengkun and Chen, Wenzhi},
  booktitle={2023 IEEE/ACM 23rd International Symposium on Cluster, Cloud and Internet Computing (CCGrid)},
  pages={13--24},
  year={2023},
  organization={IEEE}
}

@inproceedings{zhang2015nvmmu,
  title={Nvmmu: A non-volatile memory management unit for heterogeneous gpu-ssd architectures},
  author={Zhang, Jie and Donofrio, David and Shalf, John and Kandemir, Mahmut T and Jung, Myoungsoo},
  booktitle={2015 International Conference on Parallel Architecture and Compilation (PACT)},
  pages={13--24},
  year={2015},
  organization={IEEE}
}

@inproceedings{yang2023lambda,
  title={$\{$$\lambda$-IO$\}$: A unified $\{$IO$\}$ stack for computational storage},
  author={Yang, Zhe and Lu, Youyou and Liao, Xiaojian and Chen, Youmin and Li, Junru and He, Siyu and Shu, Jiwu},
  booktitle={21st USENIX Conference on File and Storage Technologies (FAST 23)},
  pages={347--362},
  year={2023}
}

@article{bergman2019spin,
  title={SPIN: Seamless operating system integration of peer-to-peer DMA between SSDs and GPUs},
  author={Bergman, Shai and Brokhman, Tanya and Cohen, Tzachi and Silberstein, Mark},
  journal={ACM Transactions on Computer Systems (TOCS)},
  volume={36},
  number={2},
  pages={1--26},
  year={2019},
  publisher={ACM New York, NY, USA}
}

@misc{amd9654,
    author = {AMD},
    title = {EPYC™ 9654},
    howpublished = {\url{https://www.amd.com/en/products/processors/server/epyc/4th-generation-9004-and-8004-series/amd-epyc-9654.html}},
}

@misc{xilinxfpga,
    author = {XILINX},
    title = {Kintex™ UltraScale+™ FPGAs},
    howpublished = {\url{https://www.amd.com/en/products/adaptive-socs-and-fpgas/fpga/kintex-ultrascale-plus.html}},
}

@misc{imagenet100,
    author = {Hugging Face},
    title = {ImageNet-100},
    howpublished = {\url{https://huggingface.co/datasets/clane9/imagenet-100}},
}

@misc{gpt2,
    author = {OpenAI},
    title = {GPT-2},
    howpublished = {\url{https://github.com/openai/gpt-2}},
}

@inproceedings{gribbon2004novel,
  title={A novel approach to real-time bilinear interpolation},
  author={Gribbon, Kim T and Bailey, Donald G},
  booktitle={Proceedings. DELTA 2004. Second IEEE international workshop on electronic design, test and applications},
  pages={126--131},
  year={2004},
  organization={IEEE}
}

@Misc{powerstore500t,
 author = {DELL},
 howpublished = {\url{https://www.delltechnologies.com/asset/en-ca/products/storage/technical-support/dell-powerstore-gen2-spec-sheet.pdf}},
 year={2023},
 title={PowerStore 500T STORAGE ARRAY.}
}

@Misc{samsung980pro,
 author={Samsung},
 howpublished = {\url{https://www.samsung.com/us/computing/memory-storage/solid-state-drives/980-pro-pcie-4-0-nvme-ssd-1tb-mz-v8p1t0b-am/}},
 year={2020},
 title={980Pro NVMe SSD}
}

@Misc{spdk,
author = {SPDK},
howpublished = {\url{https://spdk.io}},
year={2025},
title={Storage Performance Development Kit}
}

@inproceedings{kim2023nvmevirt,
  title={$\{$NVMeVirt$\}$: A versatile software-defined virtual $\{$NVMe$\}$ device},
  author={Kim, Sang-Hoon and Shim, Jaehoon and Lee, Euidong and Jeong, Seongyeop and Kang, Ilkueon and Kim, Jin-Soo},
  booktitle={21st USENIX Conference on File and Storage Technologies (FAST 23)},
  pages={379--394},
  year={2023}
}

@misc{samsungplp,
    author = {Samsung},
    title = {Power loss protection (PLP) Protect your data against sudden power loss},
    howpublished = {\url{https://download.semiconductor.samsung.com/resources/others/Samsung_SSD_845DC_05_Power_loss_protection_PLP.pdf}},
    year={2014}
}

@inproceedings{khasymski2012use,
  title={On the use of GPUs in realizing cost-effective distributed RAID},
  author={Khasymski, Aleksandr and Rafique, M Mustafa and Butt, Ali R and Vazhkudai, Sudharshan S and Nikolopoulos, Dimitrios S},
  booktitle={2012 IEEE 20th International Symposium on Modeling, Analysis and Simulation of Computer and Telecommunication Systems},
  pages={469--478},
  year={2012},
  organization={IEEE}
}

@inproceedings {scalaafa,
    author = {Shushu Yi and Xiurui Pan and Qiao Li and Qiang Li and Chenxi Wang and Bo Mao and Myoungsoo Jung and Jie Zhang},
    title = {{ScalaAFA}: Constructing {User-Space} {All-Flash} Array Engine with Holistic Designs},
    booktitle = {2024 USENIX Annual Technical Conference (USENIX ATC 24)},
    year = {2024},
    isbn = {978-1-939133-41-0},
    address = {Santa Clara, CA},
    pages = {141--156},
    url = {https://www.usenix.org/conference/atc24/presentation/yi-shushu},
    publisher = {USENIX Association},
    month = jul
}

@inproceedings{nguyen2013lightweight,
  title={A lightweight infrastructure for graph analytics},
  author={Nguyen, Donald and Lenharth, Andrew and Pingali, Keshav},
  booktitle={Proceedings of the twenty-fourth ACM symposium on operating systems principles},
  pages={456--471},
  year={2013}
}

@article{wang2017gunrock,
  title={Gunrock: GPU graph analytics},
  author={Wang, Yangzihao and Pan, Yuechao and Davidson, Andrew and Wu, Yuduo and Yang, Carl and Wang, Leyuan and Osama, Muhammad and Yuan, Chenshan and Liu, Weitang and Riffel, Andy T and others},
  journal={ACM Transactions on Parallel Computing (TOPC)},
  volume={4},
  number={1},
  pages={1--49},
  year={2017},
  publisher={ACM New York, NY, USA}
}

@inproceedings{pan2017multi,
  title={Multi-GPU graph analytics},
  author={Pan, Yuechao and Wang, Yangzihao and Wu, Yuduo and Yang, Carl and Owens, John D},
  booktitle={2017 IEEE International Parallel and Distributed Processing Symposium (IPDPS)},
  pages={479--490},
  year={2017},
  organization={IEEE}
}

@article{resnick1997recommender,
  title={Recommender systems},
  author={Resnick, Paul and Varian, Hal R},
  journal={Communications of the ACM},
  volume={40},
  number={3},
  pages={56--58},
  year={1997},
  publisher={ACM New York, NY, USA}
}

@article{burke2011recommender,
  title={Recommender systems: An overview},
  author={Burke, Robin and Felfernig, Alexander and G{\"o}ker, Mehmet H},
  journal={Ai Magazine},
  volume={32},
  number={3},
  pages={13--18},
  year={2011}
}

@inproceedings{wang2022merlin,
  title={Merlin hugectr: Gpu-accelerated recommender system training and inference},
  author={Wang, Zehuan and Wei, Yingcan and Lee, Minseok and Langer, Matthias and Yu, Fan and Liu, Jie and Liu, Shijie and Abel, Daniel G and Guo, Xu and Dong, Jianbing and others},
  booktitle={Proceedings of the 16th ACM Conference on Recommender Systems},
  pages={534--537},
  year={2022}
}

@article{achiam2023gpt,
  title={Gpt-4 technical report},
  author={Achiam, Josh and Adler, Steven and Agarwal, Sandhini and Ahmad, Lama and Akkaya, Ilge and Aleman, Florencia Leoni and Almeida, Diogo and Altenschmidt, Janko and Altman, Sam and Anadkat, Shyamal and others},
  journal={arXiv preprint arXiv:2303.08774},
  year={2023}
}

@article{liu2024deepseek,
  title={Deepseek-v3 technical report},
  author={Liu, Aixin and Feng, Bei and Xue, Bing and Wang, Bingxuan and Wu, Bochao and Lu, Chengda and Zhao, Chenggang and Deng, Chengqi and Zhang, Chenyu and Ruan, Chong and others},
  journal={arXiv preprint arXiv:2412.19437},
  year={2024}
}

@article{team2023gemini,
  title={Gemini: a family of highly capable multimodal models},
  author={Team, Gemini and Anil, Rohan and Borgeaud, Sebastian and Alayrac, Jean-Baptiste and Yu, Jiahui and Soricut, Radu and Schalkwyk, Johan and Dai, Andrew M and Hauth, Anja and Millican, Katie and others},
  journal={arXiv preprint arXiv:2312.11805},
  year={2023}
}

@inproceedings{qiu2025geminifs,
  title={$\{$GeminiFS$\}$: A Companion File System for $\{$GPUs$\}$},
  author={Qiu, Shi and Liu, Weinan and Hu, Yifan and Yan, Jianqin and Shen, Zhirong and Yao, Xin and Chen, Renhai and Zhang, Gong and Zhang, Yiming},
  booktitle={23rd USENIX Conference on File and Storage Technologies (FAST 25)},
  pages={221--236},
  year={2025}
}

@inproceedings{gofs,
  title={Managing Scalable Direct Storage Accesses for GPUs with GoFS},
  author={Li, Shaobo and Zhou, Yirui Eric and Xue, Yuqi and Xu, Yuan and Huang, Jian},
  booktitle={Proceedings of the ACM SIGOPS 31st Symposium on Operating Systems Principles},
  pages={979--995},
  year={2025}
}

@article{du2024survey,
  title={A survey of llm datasets: From autoregressive model to ai chatbot},
  author={Du, Fei and Ma, Xin-Jian and Yang, Jing-Ru and Liu, Yi and Luo, Chao-Ran and Wang, Xue-Bin and Jiang, Hai-Ou and Jing, Xiang},
  journal={Journal of Computer Science and Technology},
  volume={39},
  number={3},
  pages={542--566},
  year={2024},
  publisher={Springer}
}

@article{zhuang2023toolqa,
  title={Toolqa: A dataset for llm question answering with external tools},
  author={Zhuang, Yuchen and Yu, Yue and Wang, Kuan and Sun, Haotian and Zhang, Chao},
  journal={Advances in Neural Information Processing Systems},
  volume={36},
  pages={50117--50143},
  year={2023}
}

@inproceedings{bam,
  title={Gpu-initiated on-demand high-throughput storage access in the bam system architecture},
  author={Qureshi, Zaid and Mailthody, Vikram Sharma and Gelado, Isaac and Min, Seungwon and Masood, Amna and Park, Jeongmin and Xiong, Jinjun and Newburn, Chris J and Vainbrand, Dmitri and Chung, I-Hsin and others},
  booktitle={Proceedings of the 28th ACM International Conference on Architectural Support for Programming Languages and Operating Systems, Volume 2},
  pages={325--339},
  year={2023}
}

@article{libnvm,
  title={Smartio: Zero-overhead device sharing through pcie networking},
  author={Markussen, Jonas and Kristiansen, Lars Bj{\o}rlykke and Halvorsen, P{\aa}l and Kielland-Gyrud, Halvor and Stensland, H{\aa}kon Kvale and Griwodz, Carsten},
  journal={ACM Transactions on Computer Systems (TOCS)},
  volume={38},
  number={1-2},
  pages={1--78},
  year={2021},
  publisher={ACM New York, NY, USA}
}

@misc{SCADA1,
    author = {NVIDIA},
    title = {Advancing Memory and Storage Architectures for Next-Gen AI Workloads},
    howpublished = {\url{https://files.futurememorystorage.com/proceedings/2025/20250807_OPSW-301-1_Mailthody-2025-08-07-15.14.33.pdf}},
    year={2025}
}

@inproceedings{sun2025scalio,
  title={Scalio: Scaling up $\{$DPU-based$\}$$\{$JBOF$\}$ Key-value Store with $\{$NVMe-oF$\}$ Target Offload},
  author={Sun, Xun and Zhang, Mingxing and Shan, Yingdi and Chen, Kang and Jiang, Jinlei and Wu, Yongwei},
  booktitle={19th USENIX Symposium on Operating Systems Design and Implementation (OSDI 25)},
  pages={449--464},
  year={2025}
}

@inproceedings{li2023rubbledb,
  title={$\{$RubbleDB$\}$:$\{$CPU-Efficient$\}$ Replication with $\{$NVMe-oF$\}$},
  author={Li, Haoyu and Jiang, Sheng and Chen, Chen and Raina, Ashwini and Zhu, Xingyu and Luo, Changxu and Cidon, Asaf},
  booktitle={2023 USENIX Annual Technical Conference (USENIX ATC 23)},
  pages={689--703},
  year={2023}
}

@misc{dsfs,
    author = {DeepSeek},
    title = {Fire-Flyer File System},
    howpublished = {\url{https://github.com/deepseek-ai/3fs}},
    year={2026}
}

@misc{gpudirect,
    author = {NVIDIA},
    title = {GPUDirect},
    howpublished = {\url{https://developer.nvidia.com/gpudirect}},
    year={2026}
}

@inproceedings{zeng2026gpu,
  title={$\{$GPU$\}$$\{$Checkpoint/Restore$\}$ Made Fast and Lightweight},
  author={Zeng, Shaoxun and Ren, Tingxu and Shu, Jiwu and Lu, Youyou},
  booktitle={24th USENIX Conference on File and Storage Technologies (FAST 26)},
  pages={239--254},
  year={2026}
}

@inproceedings{yang2024demand,
  title={On-demand and parallel checkpoint/restore for gpu applications},
  author={Yang, Yanning and Du, Dong and Song, Haitao and Xia, Yubin},
  booktitle={Proceedings of the 2024 ACM Symposium on Cloud Computing},
  pages={415--433},
  year={2024}
}

@inproceedings{lee2019gpu,
  title={Gpu snapshot: checkpoint offloading for gpu-dense systems},
  author={Lee, Kyushick and Sullivan, Michael B and Hari, Siva Kumar Sastry and Tsai, Timothy and Keckler, Stephen W and Erez, Mattan},
  booktitle={Proceedings of the ACM International Conference on Supercomputing},
  pages={171--183},
  year={2019}
}

@inproceedings{wang2025finemem,
  title={$\{$FineMem$\}$: Breaking the Allocation Overhead vs. Memory Waste Dilemma in $\{$Fine-Grained$\}$ Disaggregated Memory Management},
  author={Wang, Xiaoyang and Li, Yongkun and Wu, Kan and Zhu, Wenzhe and Li, Yuqi and Xu, Yinlong},
  booktitle={19th USENIX Symposium on Operating Systems Design and Implementation (OSDI 25)},
  pages={57--74},
  year={2025}
}

@article{cai2018efficient,
  title={Efficient distributed memory management with RDMA and caching},
  author={Cai, Qingchao and Guo, Wentian and Zhang, Hao and Agrawal, Divyakant and Chen, Gang and Ooi, Beng Chin and Tan, Kian-Lee and Teo, Yong Meng and Wang, Sheng},
  journal={Proceedings of the VLDB Endowment},
  volume={11},
  number={11},
  pages={1604--1617},
  year={2018},
  publisher={VLDB Endowment}
}

@inproceedings{qin2025mooncake,
  title={Mooncake: Trading more storage for less computation—a $\{$KVCache-centric$\}$ architecture for serving $\{$LLM$\}$ chatbot},
  author={Qin, Ruoyu and Li, Zheming and He, Weiran and Cui, Jialei and Ren, Feng and Zhang, Mingxing and Wu, Yongwei and Zheng, Weimin and Xu, Xinran},
  booktitle={23rd USENIX conference on file and storage technologies (FAST 25)},
  pages={155--170},
  year={2025}
}

@book{micheloni2010inside,
  title={Inside NAND flash memories},
  author={Micheloni, Rino and Crippa, Luca and Marelli, Alessia},
  year={2010},
  publisher={Springer Science \& Business Media}
}

@incollection{pirahandeh2015energy,
  title={Energy-aware GPU-RAID scheduling for reducing energy consumption in cloud storage systems},
  author={Pirahandeh, Mehdi and Kim, Deok-Hwan},
  booktitle={Computer Science and its Applications: Ubiquitous Information Technologies},
  pages={705--711},
  year={2015},
  publisher={Springer}
}

@inproceedings{curry2010lightweight,
  title={A lightweight, gpu-based software raid system},
  author={Curry, Matthew L and Ward, H Lee and Skjellum, Anthony and Brightwell, Ron},
  booktitle={2010 39th International Conference on Parallel Processing},
  pages={565--572},
  year={2010},
  organization={IEEE}
}

@article{wu2022understanding,
  title={Understanding and exploiting the full potential of SSD address remapping},
  author={Wu, Qiulin and Zhou, You and Wu, Fei and Jiang, Hong and Zhou, Jian and Xie, Changsheng},
  journal={IEEE Transactions on Computer-Aided Design of Integrated Circuits and Systems},
  volume={41},
  number={11},
  pages={5112--5125},
  year={2022},
  publisher={IEEE}
}

@article{shao2024distributed,
  title={Distributed graph neural network training: A survey},
  author={Shao, Yingxia and Li, Hongzheng and Gu, Xizhi and Yin, Hongbo and Li, Yawen and Miao, Xupeng and Zhang, Wentao and Cui, Bin and Chen, Lei},
  journal={ACM Computing Surveys},
  volume={56},
  number={8},
  pages={1--39},
  year={2024},
  publisher={ACM New York, NY}
}

@article{zheng2022bytegnn,
  title={ByteGNN: efficient graph neural network training at large scale},
  author={Zheng, Chenguang and Chen, Hongzhi and Cheng, Yuxuan and Song, Zhezheng and Wu, Yifan and Li, Changji and Cheng, James and Yang, Hao and Zhang, Shuai},
  journal={Proceedings of the VLDB Endowment},
  volume={15},
  number={6},
  pages={1228--1242},
  year={2022},
  publisher={VLDB Endowment}
}

@article{liu2025lmcache,
  title={Lmcache: An efficient KV cache layer for enterprise-scale LLM inference},
  author={Liu, Yuhan and Cheng, Yihua and Yao, Jiayi and An, Yuwei and Chen, Xiaokun and Feng, Shaoting and Huang, Yuyang and Shen, Samuel and Zhang, Rui and Du, Kuntai and others},
  journal={arXiv preprint arXiv:2510.09665},
  year={2025}
}
